\documentclass[journal]{IEEEtran}
\ifCLASSINFOpdf
  \usepackage[pdftex]{graphicx}
  \graphicspath{{../} {../pdf/} {../jpeg/}}
  \DeclareGraphicsExtensions{.pdf,.jpeg,.png}
\else
  \usepackage[dvips]{graphicx}
  \graphicspath{{../} {../eps/}}
  \DeclareGraphicsExtensions{.eps}
\fi
\usepackage{amsmath,amssymb}
\usepackage{array}
\usepackage{colortbl}

\usepackage{multirow}
\usepackage{booktabs}
\usepackage{lipsum}
\usepackage{bm}
\usepackage{color}
\usepackage{url}

\usepackage{footnote}
\usepackage{tabularx}
\usepackage[flushleft]{threeparttable}
\usepackage{float}
\usepackage{amsthm}
\newtheorem{remark}{Remark}
\usepackage{stfloats}

\begin{document}



\title{A Data-Driven Sparse Polynomial Chaos Expansion Method to Assess Probabilistic Total Transfer Capability for Power Systems with Renewables}


%
\pdfoutput=1
\author{
~Xiaoting~Wang,~\IEEEmembership{Student Member,~IEEE,} ~Xiaozhe~Wang,~\IEEEmembership{Senior Member,~IEEE,}
~Hao~Sheng,~\IEEEmembership{Member,~IEEE,} 
~Xi~Lin,~\IEEEmembership{Member,~IEEE}
\thanks{This work is supported by Natural Sciences and Engineering Research Council (NSERC) Discovery Grant, NSERC RGPIN-2016-04570.}
}

%
%


\markboth{Accepted by IEEE Transactions on Power Systems for future publication}%
{A Data-Driven Sparse Polynomial Chaos Expansion Method to Assess Probabilistic Total Transfer Capability for Power Systems with Renewables}

%



\maketitle


\begin{abstract}
The increasing uncertainty level caused by growing renewable energy sources (RES) and aging transmission networks poses a great challenge in the assessment of total transfer capability (TTC) and available transfer capability (ATC). 
In this paper, a novel data-driven sparse polynomial chaos expansion (DDSPCE) method is proposed for estimating the probabilistic characteristics  (e.g., mean, variance, probability distribution) of probabilistic TTC (PTTC). Specifically, the proposed method, requiring no pre-assumed probabilistic distributions of random inputs,  exploits data sets directly in estimating the PTTC. 
Besides, a sparse scheme is integrated to improve the computational  efficiency. 
Numerical studies on the modified IEEE 118-bus system demonstrate that the proposed DDSPCE method can achieve accurate estimation for the probabilistic characteristics of PTTC with a high efficiency. 
Moreover, numerical results reveal the great significance of incorporating discrete random inputs in PTTC and ATC assessment, which nevertheless was not given sufficient attention. 
\end{abstract}

\begin{IEEEkeywords}
Available transfer capability (ATC), discrete random variables, polynomial chaos expansion (PCE), total transfer capability (TTC).
\end{IEEEkeywords}

%
\IEEEpeerreviewmaketitle


\section{Introduction}
%
%
%
%

\begin{figure}[h]
\setlength{\abovecaptionskip}{0.cm}
\setlength{\belowcaptionskip}{-0.8cm}
\centering
\includegraphics[width=0.35\textwidth]{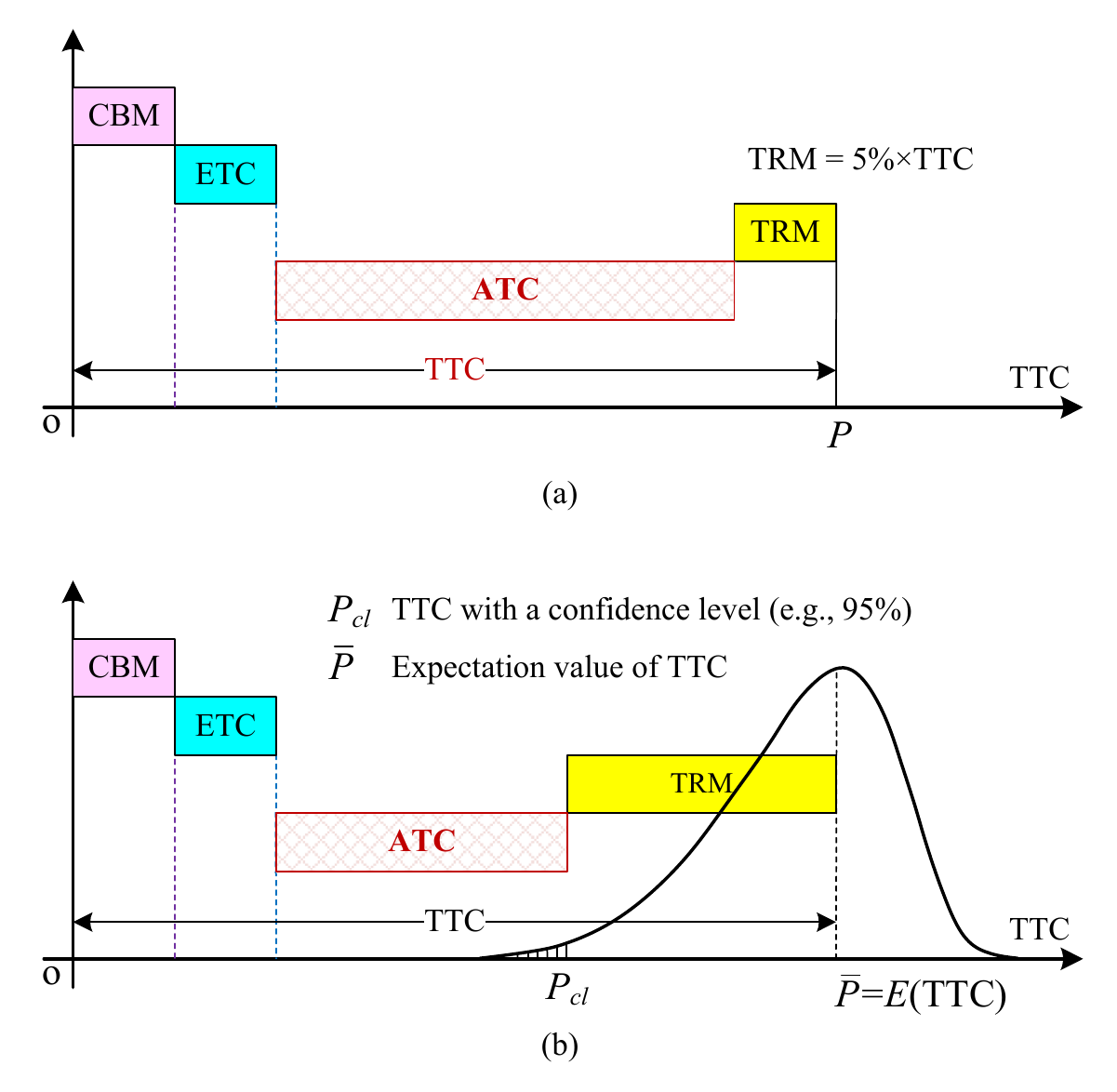}
\caption{ATC related definitions in deterministic and probabilistic frameworks. (a) Deterministic framework; (b) Probabilistic framework.}
\label{fig:prob_atc_def}
\vspace{-0.5cm}
\end{figure}

\IEEEPARstart{T}{otal} transfer capability (TTC) is defined as the amount of electric power that can be transferred over a path in the transmission network in a reliable manner while meeting all specific pre- and post-contingency system conditions. 
ATC, closely related to TTC, is a measure of the transfer capability remaining in the physical transmission network for future commercial activity over and above already committed uses \cite{NERC1996}.  Mathematically, ATC can be calculated by \cite{Li2013}:  
   $\mathrm{ATC} = \mathrm{TTC}- \mathrm{TRM}-\mathrm{ETC}-\mathrm{CBM}, $
where TRM denotes the transmission reliability margin; ETC denotes the existing transmission commitments; CBM denotes the capacity benefit margin. The relation between different quantities in the conventional deterministic framework is illustrated in Fig. \ref{fig:prob_atc_def} (a). Particularly, TRM, typically being a fixed amount or percentage (e.g., 5\% of TTC), is reserved to account for the system's various uncertainties, yet ETC is determined from the base power flow and CBM is typically determined from utility’s market model \cite{Li2013}.  

Conventionally, TTC and ATC are assessed in a deterministic framework. However, the  
growing penetration of intermittent renewable energy resource (RES) (e.g., wind farms and solar photovoltaic (PV)) and the increasing possibility of line outages due to aging transmission network  \cite{Sunday2009} have introduced a significant amount of uncertainty into power grids, which may greatly affect the TTC of a system  \cite{Hao2018}. As a result, it makes more sense to treat TTC as a random variable, i.e., probabilistic TTC (PTTC), rather than a deterministic quantity, as shown in Fig. \ref{fig:prob_atc_def} (b), while TRM needs to be carefully determined to account for the additional randomness brought about by RES and equipment failure. In other words, a probabilistic framework needs to be exploited.  
The key to evaluate ATC lies in the calculation of the statistical characteristics of PTTC, from which the TRM can also be estimated (e.g., the difference between the mean value and  the 95\% confidence interval of the TTC probability distribution). As such, the corresponding ATC evaluation is as follows: 
\begin{equation}
\label{eq:Calculate ATC}
\mathrm{ATC}=\mathbb{E}\left[\mbox{TTC}\right]-\mbox{TRM}-\mbox{ETC}-\mbox{CBM}  
\end{equation}

It is evident by comparing Fig. \ref{fig:prob_atc_def} (a) and Fig. \ref{fig:prob_atc_def} (b) that the increasing randomness from RES and other  unexpected outages may cause larger TRM and thus smaller ATC. Accurate estimation of PTTC, TRM and thus ATC is of paramount importance to ensure a secure operation of the grid while maximizing the utilization of transmission assets and economic benefits. 

In previous works, besides deterministic approaches of TTC calculation using 
continuation power flow (CPF) \cite{Chiang1995} and the repeated power flow (RPF) \cite{Ou2001}, probabilistic TTC assessment methods have also been proposed. 
Specially, Monte Carlo simulation (MCS) is the most widely used simulation-based method due to its high accuracy,  
yet it is practically unattractive due to its high computational burden 
even with advanced sampling techniques (e.g., Latin hypercube sampling (LHS) \cite{Yu2009}). 
Later on, the parametric bootstrap technique \cite{Salim2014},\cite{Othman2011} have been utilized in ATC evaluation with the distribution of random inputs known in advance. \color{black}Recently, polynomial chaos expansion (PCE), a popular uncertainty quantification method, has been applied in \cite{Haesen2009} to estimate the probabilistic load margin. The PCE method attempts to establish a statistically-equivalent model for assessing the PTTC from a small number of model evaluations. The PCE-based method has been further improved in \cite{Ni2017}\cite{Sheng2018} by using an adaptive sparse PCE  (SPCE) together with the Nataf transformation to overcome the ``curse of dimensionality'' and handle correlated random variables with different marginal distributions. 
More recently, Liu et al. \cite{Liu2019} proposed a nonparametric analytic method to calculate the PTTC. 
However, all the aforementioned methods assume that all random variables follow some parametric distribution functions (e.g., wind speed follows Weibull distribution), whereas the information about the probability distributions of random inputs may be either limited or inaccurate in practical power system applications, yet raw data of wind speed and solar radiation is more likely to be achievable  \cite{NREL2020}.  

To address this challenge, some efforts have been made to exploit data directly in assessing the PTTC. 
 Kou et al. \cite{Kou2020} proposed an interval optimization based model for ATC assessment, requiring only the boundaries of uncertain variables. 
\color{black} Li et al. \cite{Li2013} developed a point estimate method to estimate the standard deviation of PTTC directly from data. Yet the above two methods \color{black} cannot estimate the probabilistic distribution of the PTTC, which may affect the estimation accuracy of ATC.  
Zhang et al. \cite{Zhang2016}  and Xu et al. \cite{Xu2020} proposed to infer the probabilistic distributions of random inputs from the available data \cite{Zhang2016}  \cite{Xu2020} in PTTC assessment. Despite of the capability of estimating the probability distribution of PTTC, additional errors may be introduced in the procedure of approximating probabilistic distributions. 
Yan et al. \cite{Yan2018} proposed a deep learning based TTC evaluation model. However, substantial training data is required, while feature selections may be tricky. Recently, Wang et al. \cite{Wang2019} proposed a data-driven  PCE  model to solve the probabilistic power flow (PPF), which requires no probabilistic distribution assumption or approximation, yet the method may not handle discrete variables which in turn need to be considered in the PTTC assessment. Although $N-1$ contingency analysis is typically considered in PTTC calculation, it may not be sufficient to cover all unexpected events happening in the system, leading to an overestimated ATC value \cite{Audomvongseree2004}.

In this paper, we propose a novel data-driven sparse polynomial chaos expansion (DDSPCE) method that exploits available data of random inputs to assess the probabilistic characteristics (e.g., probability distribution, mean, variance) of PTTC while accounting for discrete events. To the best knowledge of the authors, the proposed DDSPCE method seems to be the \textit{first} method to account for discrete random variables in estimating the probabilistic characteristics, particularly the probability distribution, of PTTC, which will be shown to be essential and imperative. The contributions of the paper are summarized below. 
\begin{enumerate}
    \item The proposed DDSPCE, requiring no pre-assumed probability distributions of random inputs, utilizes available data directly to estimate the probabilistic characteristics (e.g., mean, variance, probability distribution) of PTTC.
    \item The proposed DDSPCE can handle a large number of mixed (both continuous and discrete) random inputs (e.g., wind speed, solar radiation, line outages occurring in the near future) in the assessment of PTTC.
    \item The proposed DDSPCE, integrating a sparse PCE scheme (i.e., the Least Angle Regression (LAR)), can produce accurate estimation for the PTTC with much less computational time compared to MCS.  
    \item  The imperative and the necessity of including discrete variables in ATC assessment is demonstrated through numerical study,  which nevertheless were not given sufficient attention.
\end{enumerate}

The rest of this paper is organised as follows. Section
II introduces the mathematical formulation of PTTC problem. Section III elaborates the DDSPCE method for estimating the PTTC. Section IV summarizes the detailed ATC computation approach using the proposed DDSPCE; Section V shows the simulation results and discussions. Section VI gives the conclusions and perspectives.


%
\section{Mathematical Formulation of Probabilistic Total Transfer Capability}
In this paper, the continuation power flow (CPF)-based method is adopted to model the PTTC problem due to its salient feature of well-conditioning at and around the critical point.  
Considering a $N$-bus transmission system, the deterministic power flow equations are expressed as 
\begin{small}
\begin{equation}
 {\bm{f}}\left( \bm{x} \right)=\begin{bmatrix}
{P_{Gi}-P_{Li}-P_{i}(\bm{x})} \\
{Q_{Gi}-Q_{Li}-Q_{i}(\bm{x})} \end{bmatrix}=0, \quad i = \{1,2,\cdots,N\} \label{eq:det power flow}
\end{equation}
\end{small}
where ${\bm{x}}=\left[\bm{\theta},\bm{V}\right]^{T}$, $\bm{\theta}$ and $\bm{V}$ denote voltage angles and magnitudes for all buses, respectively; $P_{Gi}$ and $Q_{Gi}$ denote the active and reactive power injections from the traditional generator; $P_{Li}$ and $Q_{Li}$ are the load active power and reactive power at bus $i$, respectively; ${P}_{i}\left(\bm{x}\right)$ and ${Q}_{i}\left(\bm{x}\right)$ are the total real and reactive power injections at Bus $i$. 

If we define a load-generation variation vector $\bm{b}$ to represent the direction of power transfer:
\begin{small}
\begin{equation}
\label{eq:load_gen_dir}
\bm{b}=\begin{bmatrix}
\Delta{{P}}_{G,i}-\Delta{{P}}_{L,i} \\
				-\Delta{{Q}}_{L,i} 
\end{bmatrix}, \quad i = \{1,2,\cdots,N\} 
\end{equation} 
\end{small}
where $\Delta {P}_{G,i}$, $\Delta {P}_{L,i}$, and $\Delta {Q}_{L,i}$ are the increase of active generation power, active load power, and reactive load power at bus $i$, respectively; The CPF can be formulated to study the total power that can be transferred in the specific direction $\bm{b}$:
\begin{small}
\begin{equation}
\setlength{\abovedisplayskip}{6pt}
\setlength{\belowdisplayskip}{6pt}
    \label{eq:DCPF}
    \bm{f}(\bm{x},\lambda) = \bm{f}(\bm{x})-\lambda\bm{b} = 0 
\end{equation}
\end{small}
where $\lambda \in \mathbb{R}$ is the continuation parameter indicating the transfer capability under the target direction. In the above formulation of CPF, TTC can be obtained by continuously increasing $\lambda$. 
The maximum value of $\lambda$ without violating limits (e.g., voltage limits, thermal limits, generation capacity) gives the TTC \cite{NERC1996}. 
Traditionally, TTC is calculated in this deterministic way, whereas a fixed percentage of TTC is reserved as TRM to account for a system's uncertainty. However, the increasing uncertainty level due to growing RES and aging transmission network may result in a larger TRM than expected. 
Therefore, the uncertainty needs to be carefully modelled and considered in the calculation of PTTC in order to achieve a more accurate TRM and thus ATC (see (\ref{eq:Calculate ATC})). 

\subsection{Modeling the Uncertainties}\label{section_uncertainties}
In this paper, we model wind speeds $\bm{v} (m/s)$, solar radiations $\bm{r}(W/m^2)$, load variations $\bm{P_L}$ and branch outages $\bm{\rho}$ as random variables. Note that $\bm{\rho}$ are discrete random variables, while the rest are continuous random variables. 
Particularly, wind speeds are not restricted to any probability distributions. The wind speed data can be either achieved from measurement data or generated from existing probability model. 
Once wind speed data is obtained, the wind turbine  generator’s real output power $P_{w}(v)$ can be determined  using the  wind  speed-power curve \cite{Aien2015}:
\begin{small}
\begin{equation}
\setlength{\abovedisplayskip}{6pt}
\setlength{\belowdisplayskip}{6pt}
\label{eq:wind_power}
P_{w}\left( v\right) =\left\{ 
\begin{array}{c}
0 \\ 
\frac{v-v_{in}}{v-v_{rated}}P_{rw} \\ 
P_{rw}%
\end{array}%
\right. 
\begin{array}{c}
v\leq v_{in}\text{ \rm{or} }v>v_{out} \\ 
v_{in}<v\leq v_{rated} \\ 
v_{rated}<v\leq v_{out}%
\end{array}%
\end{equation}
\end{small}
where ${v}_{in}$, ${v}_{out}$ and $ {v}_{rated}$ are the cut-in, cut-out, and rated wind speed $(m/s)$; ${P}_{rw}$ is the rated wind power $(kW)$. 

Similarly, the data of solar radiations can be either obtained from measurements or generated from existing probability model  (e.g., Beta distribution  \cite{Atwa2010}). 
The solar PV plant output power $P_{pv}(r)$ can be further calculated from the 
radiation–power curve (e.g., (4) in \cite{Sheng2019}). 
Besides, the load variation can be  represented by an assumed probability distribution (e.g., Gaussian distribution \cite{Billinton2008}) or obtained from historical data \cite{NREL2020}. 

Nevertheless, the outage of $i$th branch with a probability $q$ is modeled as a \textit{discrete} random variable, which follows the Bernoulli distribution \cite{Bae1999} with $P(\rho_i =1) = p$ and $P(\rho_i =0) = 1- p = q$. 
Again, credible historical data can also be used to describe the uncertainty of line outages. It should be noted that the equipment failure other than line outages can also be modelled and considered in the same way. 

\subsection{The Probabilistic TTC}
Let $\bm{U} = \left[\bm{v},\bm{r},\bm{P_{L},\bm{\rho}}\right]$, we can integrate the uncertainties described in Section \ref{section_uncertainties} into the deterministic power flow equations (\ref{eq:det power flow}), which becomes a set of PPF equations ${\bm{f}}\left(\bm{x},\bm{U}\right) = 0$. 
Particularly, in this paper, the wind generators are modeled as P-Q type nodes (e.g., a lagging power factor of 0.85 \cite{Aien2015}), while the solar generator is assumed to have unity power factor \cite{WECC2010}.
Similarly, the probabilistic CPF equations incorporating the random vector $\bm{U}$ can be constructed in the following compact form: 
\begin{small}
\begin{equation}
\setlength{\abovedisplayskip}{4pt}
\setlength{\belowdisplayskip}{4pt}
\label{eq:CPF}
\bm{f}\left(\bm{x},\bm{\eta}\color{black},{\lambda},\bm{U}\right)=\bm{f}\left(\bm{x},\bm{\eta}\color{black},\bm{U}\right)-\lambda \bm{b}=0 
\end{equation}
\end{small}
where $\bm{\eta}$ denotes a vector of control parameters (e.g., tap ratios of adjustable transformers). 

In this paper, we consider the voltage limits,  thermal limits, the generator active and reactive power limits as well as a credible contingency list of size  $N_c$ in the TTC calculation. 
With the preceding notation, the mathematical formulation of probabilistic TTC calculation based on the CPF method can be formed as: 

\begin{small}
\begin{subequations}
\setlength{\abovedisplayskip}{0pt}
\setlength{\belowdisplayskip}{4pt}
\label{eq:TTC_Formulation}
\begin{align}
\text{max} \quad & \lambda^{\left(\kappa\right)} \notag \\
 \text{s.t.} \quad & \bm{{f}}^{\left(\kappa\right)}\left(\bm{x},\bm{\eta},\bm{U}\right)-\lambda^{\left(\kappa\right)} \bm{b}=0 \label{eq:a} \\
 & V_{min,i}^{\left(\kappa\right)}\leq V_{i}\left(\bm{x},\bm{\eta},\lambda^{\left(\kappa\right)},\bm{U}\right) \leq V_{max,i}^{\left(\kappa\right)}, \label{eq:b} \\
 & S_{ij}^{\left(\kappa\right)}\left(\bm{x},\bm{\eta},\lambda^{\left(\kappa\right)},\bm{U}\right) \leq  S_{ij,max}^{\left(\kappa\right)}, \label{eq:c} \\
 & P_{min,i}\leq P_{Gi} \left(\bm{x},\bm{\eta},\lambda^{\left(\kappa\right)},\bm{U}\right) \leq P_{max,i}, \label{eq:d} \\
 & Q_{min,i}\leq Q_{Gi} \left(\bm{x},\bm{\eta},\lambda^{\left(\kappa\right)},\bm{U}\right) \leq Q_{max,i} \label{eq:e}
 \end{align} 
 \end{subequations}
\end{small}
where $\kappa = \{0, 1,2,\cdots, N_c\}$  with $\kappa = 0$ corresponding to the normal operation state (pre-contingency) and $\kappa = 1, 2,\cdots , N_{c}$ corresponding to the contingency cases;  
$\lambda^{\left(\kappa\right)}$ denotes the transfer capability in the $\kappa$-th case.  

TTC is the maximum value of power that can be transferred without violating any limits in both the normal operating state and in the emergency cases. I.e.,  
\begin{small}
\begin{equation}
\setlength{\abovedisplayskip}{4pt}
\setlength{\belowdisplayskip}{4pt}
    \label{eq:TTC}
    \lambda^{\text{TTC}} = \text{min}\{ \lambda^{\left(0\right)},\lambda^{\left(1\right)},\cdots,\lambda^{\left(N_{c}\right)}\}
\end{equation} 
\end{small}
It is worth noting that $\lambda$ is a random variable due to the incorporation of uncertainties, which is termed as PTTC. 
Once the mean and probability distribution of PTTC have been estimated, ATC can be obtained by  (\ref{eq:Calculate ATC}). When a large number of random variables are considered, the computational cost of calculating the probabilistic characteristics of PTTC becomes enormous. 
Besides, arbitrary distributions (e.g., continuous, discrete or mixed) and limited information (e.g., only given the raw data) of input variables also pose great challenges to the assessment of PTTC.    
%
%
\section{The Data-Driven Sparse Polynomial Chaos Expansion} \label{section:DDSPC}
In this section, the data-driven sparse polynomial chaos expansion (DDSPCE) method for PTTC assessment is elaborated in details. Unlike the conventional generalized PCE method (gPCE) \cite{Xiu2002} that requires detailed probability distributions of all random inputs, the DDSPCE method builds the orthogonal polynomial bases for mixed (both continuous and discrete) random inputs purely from data, or more specifically, moments. 
In addition, a sparse PC scheme ( e.g., LAR and modified truncation scheme) is exploited to reduce the computational effort while ensuring the accuracy of the estimation for the probabilistic characteristics of PTTC. 
\subsection{ Data-driven Polynomial Chaos Expansion} 
Consider a multi-dimensional  independent random variables $\bm{\zeta} = \left[\zeta_1,\zeta_2,\cdots, \zeta_{\mathcal{N}}\right]$, the joint cumulative distribution function  of $\bm{\zeta}$ can be represented as $\Gamma(\bm{\zeta}) = \prod_{i = 1}^{\mathcal{N}} \Gamma_{\zeta_i}(\zeta_i)$, where $\Gamma_{\zeta_i}(\zeta_i)$ is the marginal cumulative distribution function (CDF) of each individual $\zeta_i$.  
Let $Y = g(\bm{\zeta})$ be the stochastic model under consideration, 
where $Y$ is the stochastic response of interest 
(e.g. the PTTC $\lambda^{\text{TTC}}$ in (\ref{eq:TTC})),  
then the stochastic response $Y$ can be  approximated by a multi-dimensional PCE model of order $H$ \cite{Oladyshkin2012}:
\begin{small}
\begin{equation}
\setlength{\abovedisplayskip}{3pt}
\setlength{\belowdisplayskip}{3pt}
\label{eq:PCE}
    \begin{aligned}
    Y\approx \hat{Y} = \sum_{k =1}^{M}c_k \Phi_{k}(\zeta_1,\zeta_2,\cdots,\zeta_{\mathcal{N}})
    \end{aligned}
\end{equation}
\end{small}
where $M$ is the number of terms included in the expansion $\hat{Y}$. When $M$ goes to infinity, (\ref{eq:PCE}) converges in the $\mathcal{L}^2$ sense, 
whereas in practical application, the polynomial basis is typically truncated by having 
$M = (H+{\mathcal{N}})!/(H!{\mathcal{N}}!)$. Besides, 
$c_k$ refer to the unknown PCE coefficients that need to be solved; 
$\Phi_{k}(\zeta_1,\cdots,\zeta_{\mathcal{N}})$ is a set of multivariate orthogonal polynomials with respect to $\Gamma(\bm{\zeta})$. 
$\Phi_k(\zeta_1,\cdots,\zeta_{\mathcal{N}})$ can be produced by the full tensor product of one-dimensional orthogonal polynomial basis $P_i{(\zeta_i)}$. 
\begin{small}
\begin{eqnarray}
\setlength{\abovedisplayskip}{0pt}
\setlength{\belowdisplayskip}{0pt}
\label{eq:Poly_tensor}
\Phi_k(\zeta_1,\cdots,\zeta_{\mathcal{N}}) &=& \prod_{i=1}^{\mathcal{{N}}}P_i^{(\alpha_{i}^{k})}(\zeta_i)\\
    \sum_{i = 1}^{\mathcal{{N}}}\alpha_{i}^{k} \leq H, &&\quad k = \{1,2,\cdots,M\} \nonumber
\end{eqnarray}
\end{small}
where $\alpha_{i}^{k}$ correspond to the index of $k$-th expansion term of the $i$-th univariate polynomial basis, i.e.,  $\alpha_{i}^{k}$ is the degree of the univariate polynomial basis for random input $\zeta_i$ on the $k$-th expansion term. 
Note that the random variables $\bm{\zeta}$ have to be independent in the PCE method.
However, it is not uncommon that random inputs are correlated in practice (e.g. $\bm{U}$ in (\ref{eq:CPF}) is dependent in physical space). In this paper, we consider random inputs with linear correlation (spatial correlation) \color{black} in this study, while the linear correlation is removed 
by applying the principal component analysis (PCA) technique \cite{Ilin2010}. I.e.,  $\bm{\zeta} =\bm{T}_{\mathrm{pca}}({\bm{U}})$. Nevertheless, 
other techniques such as the Karhunen-Lo\'{e}ve Expansions 
\cite{Safta2016} and the maximal information coefficients calculation can also be used to remove linear correlation \cite{Liu2019}. Besides, nonlinear correlation among the random inputs may exist \cite{Konstantelos2019} in practice, requiring  other advanced techniques (e.g., kernel PCA \cite{Varon2015}, serial nonlinear transformation and the parallel nonlinear transformation \cite{Ma2018}) for decorrelation, which may be considered in our future work. 

In the conventional gPCE \cite{Xiu2002}, the choice of univariate orthogonal polynomial basis depends on the probability distribution of individual continuous random variable $\zeta_i$. However, as discussed before, data sets rather than probabilistic distributions of random inputs (e.g., wind speed) may be easier to acquire in real-world power systems. Therefore, we intend to exploit the data-driven PCE method \cite{Oladyshkin2012} 
that can build orthonormal polynomial basis described in (\ref{eq:Poly_tensor}) purely from moments that can be estimated from data. 
Therefore, $\zeta_i$ is no longer confined by pre-assumed probability distribution and 
can be continuous, discrete, described by raw data sets, or only by a limited number of moments. Certainly, it can still be described by an arbitrary probability distribution. 
\subsection{Moment-Based Polynomials}\label{section_momentbasedpoly}
The univariate orthogonal polynomial basis $P_i^{(\alpha_{i}^{k})}(\zeta_i)$ in (\ref{eq:Poly_tensor}) for the $i$-th dimensional input $\zeta_i$, $i = \{1,\cdots,{\mathcal{{N}}}\}$, is defined as below: 
\begin{small}
\begin{equation}
\setlength{\abovedisplayskip}{4pt}
\setlength{\belowdisplayskip}{4pt}
    \label{eq:Poly_expansion}
    P_i^{(l)}(\zeta_i) = \sum_{k=0}^{l} p_{k,i}^{\left(l\right)}(\zeta_i)^{k} 
\end{equation}
\end{small}
where $ l = \{0,\cdots,H\}$.  
Besides, $\alpha_{i}^{k}$ is substituted by $l$ for simplicity. $p_{k,i}^{\left(l\right)}$ is the unknown coefficient of $P_i^{(l)}(\zeta_i)$ in the $k$-th degree. 

In order to establish the polynomial basis that satisfies the orthogonality, we first demand the coefficients of leading terms for all polynomials  to be $1$, i.e., 
\begin{equation}
\setlength{\abovedisplayskip}{4pt}
\setlength{\belowdisplayskip}{4pt}
\label{eq:leading_coefficient}
p_{l,i}^{(l)} = 1, \quad \forall l 
\end{equation}
Since the formulation of polynomial basis $ P_i^{(l)}(\zeta_i) $ for each input $\zeta_i$ is the same, the formulation of polynomial basis discussed below will be applied to any individual random input $\zeta_i$. Considering the polynomial base 
must satisfy the orthogonal condition:  
\begin{equation}
\setlength{\abovedisplayskip}{4pt}
\setlength{\belowdisplayskip}{4pt}
  \label{eq:Poly_basis_Orthogonal}  
  \int_{\Omega} P_{i}^{(m)}({\zeta_i})P_{i}^{(l)}({\zeta_i})d\Gamma({\zeta_i}) = 0 \qquad \forall m \neq l
\end{equation}
where $m,l = \{0,1,\cdots,H\}$. When $l = 0$, by (\ref{eq:leading_coefficient}), we have $P_{i}^{(0)} = p_{0,i}^{(0)} = 1$. This procedure can be continued in recursive way. 
Specifically, if we consider the orthogonality conditions of polynomial $P_i^{(l)}(\zeta_i)$ with all lower-order polynomials, they can be described as below:
\begin{small}
\begin{equation}
\setlength{\abovedisplayskip}{4pt}
\setlength{\belowdisplayskip}{4pt}
\label{eq:Orthogonal_p_l}
\begin{aligned}
	& \int_{\zeta_i \in \Omega} p_{0,i}^{(0)}\left[\sum_{k = 0}^{l} p_{k,i}^{(l)}\zeta_{i}^{k} \right]d \Gamma(\zeta_i) = 0  \\ 
	&\int_{\zeta_i \in \Omega} \left[\sum_{k = 0}^{1} p_{k,i}^{(1)}\zeta^{k}\right] \left[\sum_{k = 0}^{l} p_{k,i}^{(l)}\zeta_{i}^{k} \right]d \Gamma(\zeta_i) = 0 \\ 
	& \quad  \vdots \\	
	& \int_{\zeta_i \in \Omega} \left[\sum_{k = 0}^{l-1} p_{k,i}^{(l-1)}\zeta_{i}^{k}\right] \left[\sum_{k = 0}^{l} p_{k,i}^{(l)}\zeta_{i}^{k} \right]d \Gamma(\zeta_i) = 0  \\
	&p_{l,i}^{(l)} = 1\\
\end{aligned}
\end{equation}
\end{small}
Then we substitute $p_{0,i}^{(0)}=1$ to the first equation of (\ref{eq:Orthogonal_p_l}) with (\ref{eq:leading_coefficient}), we get $p_{1,i}^{(1)}=1$. If we continue the procedure by substituting the first and second equation to the third equation, and so forth, as well as enforcing the condition (\ref{eq:leading_coefficient}), the system of equations in (\ref{eq:Orthogonal_p_l}) can be represented as: 
\begin{small}
\begin{equation}
\setlength{\abovedisplayskip}{4pt}
\setlength{\belowdisplayskip}{4pt}
\label{eq:Orthogonal_p}
\begin{aligned}
	& \int_{\zeta_i \in \Omega} \sum_{k = 0}^{l} p_{k,i}^{(l)}\zeta_{i}^{k} d \Gamma(\zeta_i) = 0  \\ 
	&\int_{\zeta_i \in \Omega} \sum_{k = 0}^{l} p_{k,i}^{(l)}\zeta_{i}^{k+1} d \Gamma(\zeta_i) = 0 \\ 
	& \quad  \vdots \\	
	& \int_{\zeta_i \in \Omega}  \sum_{k = 0}^{l} p_{k,i}^{(l)}\zeta_{i}^{k+{l}-1} d \Gamma(\zeta_i) = 0  \\
	& p_{l,i}^{(l)} = 1
\end{aligned}
\end{equation}
\noindent Besides, recall that the $k$-th raw moment of $\zeta_i$ is defined as below:
\begin{equation}
\setlength{\abovedisplayskip}{4pt}
\setlength{\belowdisplayskip}{4pt}
\label{eq:moment_l}
    \mu_{k,i} = \int_{\zeta_i \in \Omega} \zeta_{i}^{k}d\Gamma(\zeta_i)
\end{equation}
\end{small}
As a result, the set of equations in (\ref{eq:Orthogonal_p}) can be described by the moments of $\zeta_i$ and put into the matrix form:
\begin{small}
\begin{equation}
\setlength{\abovedisplayskip}{4pt}
\setlength{\belowdisplayskip}{4pt}
\label{eq:matrix_coefficients}
	\left[\begin{array}{cccc}
	\mu_{0,i} & \mu_{1,i} &\ldots & \mu_{l,i} \\
	\mu_{1,i} & \mu_{2,i} &\ldots & \mu_{l+1,i} \\
	\vdots & \vdots & \vdots & \vdots \\
	\mu_{l-1,i} & \mu_{l,i} & \ldots & \mu_{2l-1,i} \\
	0 & 0 & \ldots & 1
	\end{array}\right] \left[\begin{array}{c}
	p_{0,i}^{(l)} \\ p_{1,i}^{(l)} \\\vdots \\ p_{l-1,i}^{(l)} \\ p_{l,i}^{(l)}
	\end{array}\right] = \left[\begin{array}{c}
	0\\0\\ \vdots \\ 0\\1
\end{array}\right]
\end{equation}
\end{small}
where the $j$th moment of the $i$th random input $\mu_{j,i}, \quad j = \{0,1,\cdots,2{l}-1\} $ and $i = \{1,\cdots,{\mathcal{{N}}}\}$ can be approximated from the samples by
$\mu_{j,i} = \frac{1}{M_p}\sum_{m=1}^{M_p} \zeta_{m,i}^{j}$, 
where $M_p$ is the sample size; $\zeta_{m,i}$ are the sample points of an arbitrary input random variable $\zeta_i$. Once the raw data set or distributions of $\zeta_i$ are given, the raw moment of $\zeta_i$ can be obtained. 

It's clear from (\ref{eq:matrix_coefficients}) that the availability of any $l$-th order  polynomial chaos expansion requires the existence of moments $\mu_0,\mu_1,\cdots,\mu_{2{l}-1}$. 
The orthogonal polynomial basis $P_{i}^{(l)}(\zeta_i)$ for each random input can be formulated from (\ref{eq:Poly_expansion}) after solving (\ref{eq:matrix_coefficients}) for $p_{k,i}^{(l)}$.  Note that $P_i^{(l)}(\zeta_i)$ can be used directly for analysis, yet normalized basis gives more useful properties. Therefore, a normalization procedure is conducted \cite{Oladyshkin2012}: 
\begin{small}
\begin{equation}
\setlength{\abovedisplayskip}{4pt}
\setlength{\belowdisplayskip}{4pt}
    \label{eq:normalization}
    \psi_i^{(l)}(\zeta_i) = \frac{1}{\Vert P_i^{(l)}(\zeta_i)\Vert} \sum_{k=0}^{l}p_{k,i}^{(l)}\zeta_i^{k}, \quad l = \{0,\cdots,H\}
\end{equation}
\end{small}
where 
$\Vert P_i^{(l)}(\zeta_i)\Vert$ is the norm of $ P_i^{(l)}(\zeta_i)$:
\begin{small}
\begin{equation}
\setlength{\abovedisplayskip}{4pt}
\setlength{\belowdisplayskip}{4pt}
    \label{eq:norm_poly}
    \begin{aligned}
         \int_{\zeta_i \in \Omega} \left[P_i^{(l)}(\zeta_i) \right]^2 d\Gamma(\zeta_i) 
        = & \int_{\zeta_i \in \Omega} \left[\sum_{k=0}^{l}p_{k,i}^{(l)}\zeta_{i}^{k} \right]^2 d\Gamma(\zeta_i) \\
         = & \sum_{k=0}^{l}\sum_{j=0}^{l}p_{k,i}^{(l)}p_{j,i}^{(l)}\mu_{k+j,i}
    \end{aligned}
\end{equation}
\end{small}
From (\ref{eq:norm_poly}), it can be seen that at least $2l$-th order raw moments of $\zeta_i$ are needed to construct a normalized $l$-th order orthogonal polynomial basis. 
Once the one dimensional orthonormal polynomial base are constructed, the multi-dimensional orthonormal polynomials can be determined referring to (\ref{eq:Poly_tensor}), where the $P_i^{(\alpha_{i}^{k})}(\zeta_i)$ are replaced by $\psi_i^{(l)}(\zeta_i)$. 

It is worth noting that the one dimensional orthonormal basis $\{\psi_i^{(0)}(\zeta_i),\cdots, \{\psi_i^{(l)}(\zeta_i)\}$ ($l= \{0,\cdots,H\}$) can be built \textit{if and only if} the the $2l$-th order raw moments (e.g., $\mu_{0,i},\cdots,\mu_{2l,i}$) exist and the number of support points (distinct values of all sample points) of $\zeta_i$ is greater than the desired degree $H$ of basis if $\zeta_i$ is a discrete random variable or represented by a data set \cite{Oladyshkin2012}. In the calculation of PTTC, a large number of correlated mixed random variables $\bm{U}$ with sample size $M_p$ are considered, where $M_p$ is generally much larger than the desired degree $H$. As such, 
$\bm{\zeta}$ transformed from $\bm{U}$ through PCA typically satisfies the above condition for a specified degree $H$. 

Once the orthogonal polynomial basis is built from the data, we need to calculate the expansion coefficients $c_k$ $k =1,\cdots,M$ to build the surrogate PCE model (\ref{eq:PCE}).  
\vspace{-0.3cm}
\subsection{Calculation of Expansion Coefficients} \label{section_PCEcoefficients}
The least-square regression method is used to calculate the expansion coefficients $c_k$. Specifically, for a set of sample-response pairs $\left[ \bm{\zeta_p}, \bm{Y_p}\right]$ with $\bm{\zeta} = \left\{{\bm{\zeta}^{(1)}},\bm{\zeta}^{(2)}\cdots,\bm{\zeta}^{(M_p)}\right\}$ and $\bm{Y} = \left\{Y^{(1)},Y^{(2)},\cdots,Y^{(M_p)} \right\}$, the expansion coefficients $c_k$ can be computed by minimizing the sum of the squared residuals: 
    \begin{small}
    \begin{equation}
    \setlength{\abovedisplayskip}{4pt}
    \setlength{\belowdisplayskip}{4pt}
    \label{eq:cost_fun}
    \begin{aligned}
        J(\bm{C}) & = \sum_{m =1}^{M_p}\left[ Y^{(m)} - \sum_{k=1}^{M}c_k \Phi_k(\bm{\zeta}^{(m)})\right]^2 \\
        & = \left(\bm{Y} - \bm{\Psi}\bm{C}\right)^{{T}}\left(\bm{Y} - \bm{\Psi}\bm{C}\right)
    \end{aligned}
    \end{equation}
    \end{small}
where $\bm{C} = [c_1,c_2,\dots,c_M]^{{T}}$ and $\bm{\Psi}_{ij} = \Phi_{j}(\bm{\zeta}^{(i)})$ with $i =\{1,2,\cdots,M_p\}$ and $j = \{1,2,\cdots,M\}$.
By taking the derivative of (\ref{eq:cost_fun}), the ordinary least-square solution of (\ref{eq:cost_fun}) can be obtained as follows:
\begin{small}
\begin{equation}
\setlength{\abovedisplayskip}{4pt}
\setlength{\belowdisplayskip}{4pt}
    \label{eq:ols_soln}
    \bm{\hat{C}} =  \left(\bm{\Psi}^{{T}}\bm{\Psi}\right)^{-1}\bm{\Psi}^{{T}}\bm{Y}
\end{equation} 
\end{small}
\vspace{-0.8cm}
\subsection{A Sparse PCE Scheme: Least Angle Regression (LAR)}\label{section_LAR}
\noindent\textbf{Modified Truncation Scheme:}
The number of terms in the multi-dimensional orthonormal polynomials $\Phi_{k}(\bm{\zeta})$ will grows exponentially with the number of random inputs. 
Since the low-order input variables are sufficient in most practical problems \cite{Marelli2018}, 
the hyperbolic (or $q$-norm) truncation scheme is employed to save computational cost \cite{Marelli2018}:
\begin{small}
\begin{equation}
\setlength{\abovedisplayskip}{4pt}
\setlength{\belowdisplayskip}{4pt}
\label{eq:multi_index_modify}
    \left(\sum_{i=1}^{\mathcal{{N}}}\left(\alpha_{i}^{k}\right)^q\right)^{\frac{1}{q}} \leq H, \quad k = 1,\cdots, M
\end{equation}
\end{small}
where $q \in (0,1)$. $k$ denotes the number of expansion term. As such, a sparser PCE can be produced compared to the standard truncation in (\ref{eq:Poly_tensor}).  

\noindent \textbf{The LAR Algorithm:} LAR is an efficient linear regression tool for variable selection, which aims to select the predictors (e.g., the polynomial base $\bm{\Phi_{k}}$ in (\ref{eq:PCE})) that are most relevant to the model response ${Y}$ (e.g., the PTTC) 
from a possible set with large candidates. 
See \cite{Marelli2018} for the detailed steps of LAR procedure. Following a number of iterations, LAR eventually builds a sparse PCE approximation with reduced expansion terms. 

Besides, the corrected leave-one-out cross-validation error ($e_{cloo}$) has been adopted as stop criteria \cite{Marelli2018} to reduce the computational burden. The $e_{cloo}$ is calculated by:
\begin{small}
\begin{equation}
\setlength{\abovedisplayskip}{4pt}
\setlength{\belowdisplayskip}{4pt}
\label{eq:ecloo}
 e_{cloo} = T(M,M_p) e_{loo} 
\end{equation}
\end{small}
with
\begin{small}
\begin{equation}
\setlength{\abovedisplayskip}{4pt}
\setlength{\belowdisplayskip}{4pt}
\label{eq:eloo}
\left\{\begin{aligned}
 & e_{loo} = \frac{\sum_{m =1}^{M_p}\left[ \frac{Y^{(m)} - \hat{Y}^{(m)})}{{1-h_i}}\right]^2}{\sum_{m =1}^{M_p}\left[ Y^{(m)}-\hat{\mu}_Y\right]^2} \\
 & T(M,M_p) = \frac{M_p}{M_p -M}\left(1+{\mbox{tr}\left[\left(\bm{\Psi}^{T}{\bm{\Psi}}\right)^{-1}\right]}\right)
\end{aligned}\right.
\end{equation}
\end{small}
where $\bm{h} = \mbox{diag}\left(\bm{\Psi}\left(\bm{\Psi}^{T}{\bm{\Psi}}\right)^{-1}\bm{\Psi}^{T}\right)$ and $h_i$ is its $i$th element; $\hat{\mu}_Y = \frac{1}{M_p}\sum_{m=1}^{M_p}Y^{(m)}$ is the sample average; $T(M,M_p)$ is the correction factor which will increase with the number of expansion terms $M$ and achieve $1$ when the size of samples $M_p \rightarrow{\infty}$. 

The sparse PCE scheme with such stop criteria has the following notable properties compared with the conventional leave-one-out error $e_{loo}$-based stop criteria  \cite{Marelli2018}: i) it is more computation effective since it requires to run LAR procedure only once for all $M_p$ samples, while the conventional one requires $M_p+1$ times; ii) it is more effective in overcoming the over-fitting problem; iii) it works well with a small training sample size in the context of PCE.  


\section{ATC Computation Approach}\label{section_algorithm}

In this section, a detail description of the proposed algorithm in ATC evaluation is provided. Particularly, the developed DDSPCE is used to estimate the probabilistic characteristics of PTTC, from which ATC with certain confidence level is assessed by (\ref{eq:Calculate ATC}).

\noindent \textbf{Step 1.} Input the network data, the contingency list and the transaction of interest. Input $M_p$ samples of ${\mathcal{{N}}}$ random inputs (e.g., wind speeds, solar radiations, active load power and the status of branches) $\bm{U_p} = \left(\bm{U}^{(1)},\cdots,\bm{U}^{(M_p)} \right)  \in \mathbb{R}^{M_p \times {\mathcal{{N}}}}$ obtained either through raw data or assumed probabilistic distributions.

\noindent \textbf{Step 2.} Build the load-generation vector $\bm{b}$ (see (\ref{eq:load_gen_dir})) according to the transaction under study and evaluate the PTTC $\bm{Y_p} = \left(Y^{(1)},\cdots,Y^{(M_p)}\right)$ that correspond to $\bm{U_p}$ by solving (\ref{eq:TTC_Formulation}). 

\noindent \textbf{Step 3.} Decorrelate random variables $\bm{U_p}$ by applying the PCA technique, which transform $\bm{U_p}$ to the independent samples $\bm{\zeta_p} = \left(\bm{\zeta}^{(1)},\cdots,\bm{\zeta}^{(M_p)}\right)$. 
Pass the data set $\left[ \bm{\zeta_p}, \bm{Y_p}\right]$ to \textbf{Step 4}. 

\noindent \textbf{Step 4.} Apply the moment-based method discussed in Section \ref{section_momentbasedpoly} to construct the one-dimensional orthonormal polynomial basis $\psi^{(l)}(\zeta_i)$ for each $\zeta_i$. More specifically, Specify the PCE degree $H$. For each random variable $\zeta_i$,  $l=\{0,1,\cdots,H\}$. 
\begin{description}
\item [4a)] Calculate $0$ to $2{l}$-th moments;  
\item [4b)] Solve (\ref{eq:matrix_coefficients}) for the polynomial coefficients: $p_{k,i}^{\left(l\right)}$, $ k =0,\cdots,l$;   
\item [4c)] Form the orthogonal univariate polynomial basis $P_{i}^{(l)}(\zeta_i)$ 
by (\ref{eq:Poly_expansion}).
\item [4d)] Normalize $P_{i}^{(l)}(\zeta_i)$ by (\ref{eq:normalization}) to get the orthonormal polynomials $\psi_{i}^{(l)}(\zeta_i)$.  
\end{description}
\noindent \textbf{Step 5.}  Apply the algorithms discussed in Section \ref{section_PCEcoefficients}, D to build the data-driven sparse PCE model (\ref{eq:PCE}) for calculating the PTTC.  
\begin{description}
    \item [5a)] Truncate the index $l$ using (\ref{eq:multi_index_modify}) and construct the corresponding multi-dimensional polynomial base of degree $H$ to build the matrix $\bm{\Psi}$ in (\ref{eq:cost_fun}).
    \item [5b)] Implement the LAR algorithm to find the optimal sparse polynomial base and compute the expansion coefficients $c_k$ by (\ref{eq:ols_soln}). 
    \item [5c)] Compute the corrected leave-one-out error $e_{cloo}$ according to (\ref{eq:ecloo}) and (\ref{eq:eloo}). Go to \textbf{Step 6}. 
\end{description}
\noindent \textbf{Step 6.} If the data-driven sparse PCE model 
has reached the prescribed accuracy (e.g., $e_{cloo} < e_{stop}$)\color{black}, go to \textbf{Step 7}. Otherwise, enlarge the size of training set by $\Delta M_p$ (e.g. $\bm{U_{\Delta p}}$) and calculate $\bm{Y_{\Delta p}}$ by solving (\ref{eq:TTC_Formulation}), then let $M_p\leftarrow M_p+\Delta p$, $\bm{U_p}\leftarrow (\bm{U_p},\bm{U_{\Delta p}}), \bm{Y_p} \leftarrow (\bm{Y_p},\bm{Y_{\Delta p}})$ and go back to \textbf{Step 3}.

\noindent \textbf{Step 7.} Once the PCE model is built, acquire additional $M_s$ sample points of $\bm{U}$ from raw data or their assumed probability distributions. Apply the PCA technique to transform all samples to $\bm{\zeta_s} = \left(\bm{\zeta}^{(1)},\cdots,\bm{\zeta}^{(M_s)}\right)$ and compute the PTTC $\bm{Y_s} = \left({Y}^{(1)},\cdots,{Y}^{(M_s)}\right)$ using the established PCE model (\ref{eq:PCE}).  

\noindent \textbf{Step 8.} Compute the statistics of the PTTC, e.g.,  mean value, standard deviation, probability density function (PDF) and CDF. 

\noindent \textbf{Step 9.} Determine the TRM and the corresponding ATC value based on a desired confidential level $P_{cl}\%$, i.e., $P(ATC_{\mathrm{actual}} \geq (\mu_{TTC}-TRM)) = P_{cl}\%$ and generate result report.
\vspace{-0.15cm}
\begin{remark} The number of $M_s$ in \textbf{Step 7} is much larger than $M_p$ in \textbf{Step  2}, indicating that the computational effort for evaluating PTTC evaluation through (\ref{eq:TTC_Formulation}) is greatly reduced in DDSPCE compared to MCS. It is worth noting that most of the computational cost of DDSPCE lies in \textbf{Step 2}. The DDSPCE method is much more computationally efficient compared to MCS 
since $M_p \ll M_s$. 
\end{remark}
\vspace{-0.3cm}
\begin{remark}
The empirical value of the training sample size ($M_p$ in \textbf{Step 3}-\textbf{5}) required by the DDSPCE method is about $5$ times the number of random inputs (e.g., $M_p \approx 5{\mathcal{N}}$). So if the initial value of $M_p$ starts around $5{\mathcal{N}}$, less iterations may be needed in the implementation of the proposed algorithm. 
\end{remark}
\vspace{-0.3cm}
\begin{remark} Control devices (e.g., adjustable transformers, switchable shunts) and $N-K$ contingency can be readily incorporated in the PTTC formulation and ATC assessment. Numerical examples are given in Section \ref{section_scenario3} for illustration. 
\end{remark}
%

\section{Numerical Studies and Discussions}
In this section, we applied the proposed DDSPCE method to assess the PTTC of the modified IEEE 118-bus system \cite{Zimmerman2011}. Particularly, three \color{black}scenarios are considered. The first scenario has only continuous random variables (i.e., wind speed $\bm{v}$, solar radiation $\bm{r}$ and load variation $\bm{P_L}$), while the second \color{black}scenario considers both continuous random variables and discrete random variables (i.e., line outages). It will be shown that incorporating discrete variables in PTTC assessment is essential as they will greatly affect the statistics of PTTC and reduce the final ATC value. In the third scenario,  $N-2$ contingency and adjustable transformers are considered, respectively, to show that they can be readily incorporated into the formulation of PTTC and ATC assessment.

In this paper, the probabilistic data of all random variables are generated from assumed probability distributions (e.g., see Section \ref{section_uncertainties}), yet only the data is exploited and the information of probability distributions are not used. The readers can find the parameter details in  {\url{https://github.com/TxiaoWang/DDSPC-TTC}}. 
In practical applications, if sufficient credible raw data is known, the PTTC can be directly assessed by the proposed method. 

All simulations have been conducted on the MATLAB R2018b on a PC equipped with Intel Core i7-8700 (3.20GHz), 16GB RAM. The deterministic simulation tool used to calculate exact TTC values is Voltage Security Assessment Tool (VSAT), a core toolset of DSATools; Toolbox UQLab is adopted to build the DDSPCE scheme \cite{Marelli2018}. 
\vspace{-0.2cm}
\subsection{Scenario 1: With Only Continuous Random Inputs}
The IEEE 118-bus system \cite{Zimmerman2011} includes 19 generators, 35 synchronous condensers, 177 transmission lines, and 91 loads. The total load of this network is 4242 MW and 1438 MVar. In this Scenario, 111 continuous random variables are incorporated into the IEEE 118-bus system, 
which include wind speeds $\bm{v}$, solar radiations $\bm{r}$ and load power $\bm{P_L}$. To be specific, there are 6 wind farms connected to bus \{10, 25, 26, 49, 65, 66\}, six solar PV plants connected to bus \{12, 59, 61, 80, 89, 100\} and 99 probabilistic loads. The new transaction under study is the power transfer from the generators at Bus \{87, 89, 111\} to the loads at Bus \{88, 90, 91, 92, 103\}. The contingency list contains five $N - 1$ outages, which are \{L88-89, L7-12, L13-15, L49-54, L91-92\}. 

Firstly, the DSATools/VSAT solver is  applied to calculate the total transfer capability of the deterministic system (no uncertainty). The resulting deterministic TTC is 139.9 MW \color{black}considering voltage limits, thermal limits, generation capacity,  and stability limit. See Table \ref{tab:DTTC} for the deterministic TTC in normal and contingency cases. 
\begin{table}[H]
\vspace{-0.3cm}
\setlength{\abovecaptionskip}{0.cm}
\renewcommand{\arraystretch}{1.3}
\centering
\caption{The deterministic TTC in normal and contingency cases}
\label{tab:DTTC}
\begin{tabular}{c|c|c|c}
\hline
Case
No. &    Outage Facility & Violation type           & TTC(MW)  \\ 
\hline
0        &  Base case       &  Max
 generation violation &  299      \\ 
\hline
1        & L88-89             &  Voltage
violation         & 139.9    \\
\hline
2        &  L7-12              &  Max generation violation &  299      \\
\hline
3        &  L13-15             &  Max generation violation &  299      \\
\hline
4        &  L49-54             &  Max generation violation &  299      \\
\hline
5        &  L91-92             &  Max generation violation &  299    \\
\hline
\end{tabular}
\end{table}

Next, the proposed DDSPCE method is applied to estimate the probabilistic characteristics of the PTTC (e.g. mean, standard deviation, and PDF/CDF).  Particularly, to show how to select the order of PCE model $H$ and the sample size $M_p$, Table \ref{tab:TTC_comp_samplesize_order} presents comparisons of the corrected leave-one-out-error $e_{cloo}$, the estimated mean value $\mu$ and the standard deviation $\sigma$ of PTTC, and the normalized estimation errors for different combinations. As can be seen that the $e_{cloo}$ grows when $H$ increases from 2 to 3, thus $H=2$ is selected for the PCE model. 
In addition, the comparison of estimation accuracy using different sample size $M_p$ confirms \textbf{Remark 2} that the empirical value of $M_p$ required by the DDSPCE method is about 5 times the number of random inputs for sufficiently good estimation accuracy. Hence, in the cases thereafter, 
$H =2$ and $M_p =556$. 
\begin{table*}[]
\setlength{\abovecaptionskip}{0.cm}
\renewcommand{\arraystretch}{1.3}
\caption{Comparison of estimation accuracy of the DDSPCE method for different sample sizes $M_p$ and model orders $H$}
\label{tab:TTC_comp_samplesize_order}
\centering
\begin{tabular}{c|c|c|c|c|c|c|c|c|c|c}
\hline
\multicolumn{1}{c|}{\multirow{2}{*}{$M_p$}} & \multicolumn{5}{c|}{ $H=2$}                                                                                                                                              & \multicolumn{5}{c}{ $H=3$}                                                                                                                                              \\ \cline{2-11} 
\multicolumn{1}{c|}{}                        & \multicolumn{1}{c|}{$e_{cloo}$} & $\mu$    & \multicolumn{1}{c|}{$\frac{\Delta\mu}{\mu_{MCS}}$\%} & $\sigma$ & \multicolumn{1}{c|}{$\frac{\Delta\sigma}{\sigma_{MCS}}$\%} & \multicolumn{1}{c|}{$e_{cloo}$} & $\mu$    & \multicolumn{1}{c|}{$\frac{\Delta\mu}{\mu_{MCS}}$\%} & $\sigma$ & \multicolumn{1}{c}{$\frac{\Delta\sigma}{\sigma_{MCS}}$\%} \\ \hline
278                                           & 0.0264                          & 138.6281 & -1.3652                                            & 30.9066  & 4.2371                                                   & 0.1509                          & 138.6281 & -1.3652                                            & 26.3000  & -11.2994                                                 \\ \hline
417                                           & 0.0172                          & 139.4681 & -0.7675                                            & 30.2735  & 2.1018                                                   & 0.0554                          & 139.4681 & -0.7675                                            & 28.0600  & -5.3635                                                  \\ \hline
556                                           & 0.0120                          &  139.8183 &  -0.5183                                            & 29.7303  & 0.2698                                                   &  0.0153                          & 139.8183 & -0.5183                                              & 28.8977  & -2.5383                                                  \\ \hline
\end{tabular}
\vspace{-0.5cm}
\end{table*}

The comparison of  DDSPCE, SPCE \cite{Sheng2018} with respect to MCS is presented in Table \ref{tab:TTC_comp_Dif},
which includes the mean value $\mu$, the standard deviation $\sigma$ and their normalized errors in $\%$. 
It can be obviously observed that the proposed DDSPCE can provide very accurate estimation results compared to the benchmark MCS. In addition, Fig. \ref{fig:TTC_comp} presents comparisons of the estimated PDF and CDF of the PTTC (from $M_s = 10000$ samples) by the MCS, DDSPCE, and SPCE methods, respectively. As can be seen, results from these three methods are almost overlapped, indicating a good accuracy of the proposed DDSPCE method. However, unlike SPCE, the proposed DDSPCE requires no pre-assumed probability distributions of the random inputs. 
\begin{table}[H]
\setlength{\abovecaptionskip}{0.cm}
\setlength{\belowcaptionskip}{-0.5cm}
\vspace{-0.3cm}
\renewcommand{\arraystretch}{1.3}
\caption{Comparison of the estimated statistics of the overall TTC by the MCS, DDSPCE and SPCE methods}
\label{tab:TTC_comp_Dif}
\centering
\begin{tabular}{c|c|c|c}
\hline
Index & MCS & DDSPCE (proposed) & SPCE \cite{Sheng2018} \\
\hline
$\mu$ & 140.5468 & 139.8183 &  140.6716 \\
\hline 
 $\sigma$  & 29.6503  & 29.7303 & 29.4201\\ 
 \hline 
 $\frac{\Delta\mu}{\mu_{MCS}}\%$ & -- & -0.5183  & 0.0888\\ 
 \hline
 {$\frac{\Delta\sigma}{\sigma_{MCS}}\%$} & -- & 0.2698 &-0.7764 \\
 \hline
\end{tabular}
\vspace{-0.3cm}
\end{table}
\begin{figure}[]
\setlength{\abovecaptionskip}{0.cm}
\vspace{-0.1cm}
\setlength{\belowcaptionskip}{-0.5cm}
\centering
\includegraphics[width=0.38\textwidth]{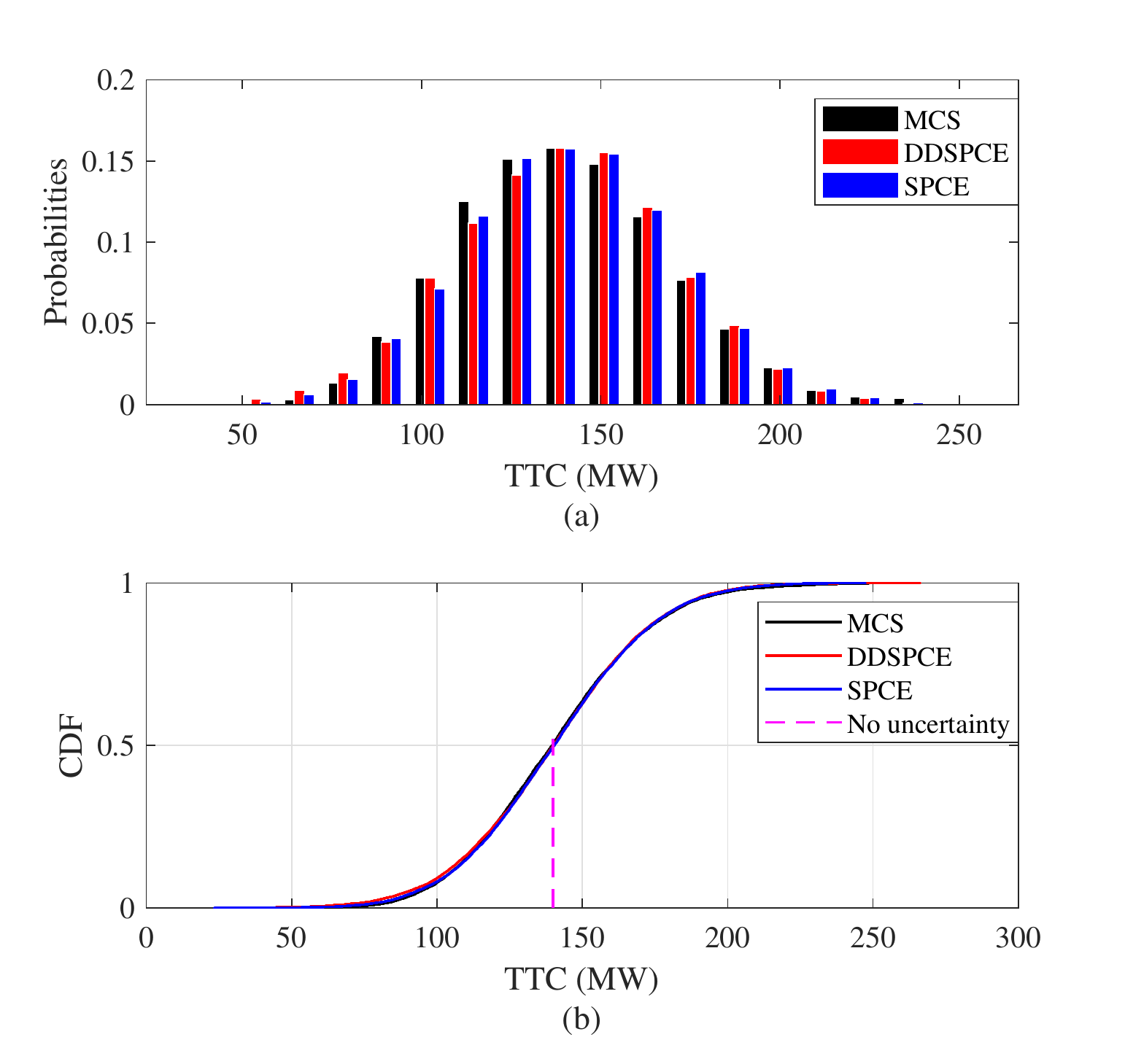}
\caption{The PDF and CDF of the PTTC calculated by the MCS, DDSPCE and the SPCE. They are almost overlapped. The TRM for 95\% confidence level is 49.7043 MW and the corresponding ATC is 90.8425 MW.} 
\label{fig:TTC_comp}
\vspace{-0.5cm}
\end{figure}

Regarding the efficiency, a comparison of computation time between DDSPCE, SPCE and MCS is presented in Table \ref{tab:IEEE118_time_comp}, in which $t_{\mathrm{ed}}$ denotes the time for generating data set $\left[ \bm{\zeta_p}, \bm{Y_p}\right]$ ($M_p = 556$); $t_{\mathrm{sc}}$ is the time for constructing the PCE model (e.g. DDSPCE and SPCE); $t_{\mathrm{es}}$ is the time for computing the PTTC of $M_s = 10000$ samples. It is clear that the time consumption of the DDSPCE is substantially smaller than that of MCS, i.e. about ${\frac{1}{18}}$ of time consumption required by MCS. Moreover, the time for constructing the DDSPCE model ($t_{\mathrm{sc}}$) is about $5$ times less than the SPCE method. 

Once the statistical characteristics of PTTC have been obtained, acceptable TRM values, i.e., the difference between the mean value of TTC and TTC value at a specified confidence level, can be determined based on the CDF of PTTC, and thus ATC. Table \ref{tab:IEEE118_confid_levels} presents the estimated TRM values according to some particular confidence levels and the corresponding ATC values. Specially, for a desired confidence level 95\%, $P(ATC_{\mathrm{actual}} \geq (\mu_{TTC}-TRM)) = 0.95$, the TRM is 49.7043 MW and the resulting ATC is 90.8425 MW.   
\begin{table}[H]
\setlength{\abovecaptionskip}{0.cm}
\vspace{-0.3cm}
\renewcommand{\arraystretch}{1.3}
\caption{Comparison of Computation Time between DDSPCE and MCS, and SPCE}
\label{tab:IEEE118_time_comp}
\centering
\begin{tabular}{c|c|c|c|c}
\hline
Method & ${t_{\mathrm{ed}}(s)}$ & ${t_{\mathrm{sc}}(s)}$ & ${t_{\mathrm{es}}(s)}$ & ${t_{\mathrm{total}}(s)}$ \\
\hline
MCS    & --  &  --    & 173484.23 & 173484.23  \\
\hline
DDSPCE  &   9368.61   &   1.09     & 0.33 \color{black}&  9370.03 \\
\hline
 SPCE   &  9368.61   &  5.46  & 0.35 \color{black} &  9374.42 \\
\hline 
\end{tabular}
\vspace{-0.5cm}
\end{table} 
\begin{table}[H]
\setlength{\abovecaptionskip}{0.cm}
\renewcommand{\arraystretch}{1.2}
\caption{The estimated TRM and resulting ATC (MW) for different confidence levels based on the DDSPCE model}
\label{tab:IEEE118_confid_levels}
\centering
\begin{tabular}{c|c|c|c}
\hline
Confid. Level & $\mathbb{E}($TTC$)$ (MW) & {TRM} (MW) & {ATC} (MW) \\ 
\hline
${99.0\%}$ & 140.5468 & 72.0892 & 68.4576   \\
\hline
${98.0\%}$ & 140.5468 & 62.4110 & 78.1358  \\
\hline
${95.0\%}$ & 140.5468 & 49.7043 & 90.8425   \\
\hline
${90.0\%}$ & 140.5468 & 38.2358 & 102.3110  \\
\hline
${80.0\%}$ & 140.5468 & 25.4402 & 115.1066 \\
\hline
\end{tabular}
\vspace{-0.4cm}
\end{table}
\subsection{Scenario 2: With Mixed Random Inputs} 

In the second scenario, we further incorporate four line outages as discrete random variables into IEEE 118-bus system to test the performance of the proposed method and illustrate the impacts of these discrete events on PTTC and ATC. The configuration for the continuous random variables, the transaction under study and $N-1$ contingency are the same as those in Scenario 1. There are totally 115 random variables including four additional line outages, which are \{L89-90, L90-91, L89-92, L92-94\}. For simplicity, we assume the probabilities $q$ of unavailability for each line are the same and independent. Particularly, two cases $q=0.1$ and $q=0.2$ are considered. 

Similarly, the deterministic TTC MW under new transaction using DSATools/VSAT is calculated as $139.9$ MW. Next, the proposed DDSPCE method is applied to evaluate the probabilistic characteristics of the PTTC for two cases ($q = 0.1$ and $q=0.2$). The order of PCE model $H$ is set as 2 for both cases. To construct the DDSPCE model (\ref{eq:PCE}), 556 simulations are required ($M_p$ in \textbf{Step 3}-\textbf{5}). 10000 samples ($M_s$ in \textbf{Step 7}) are generated to estimate the probabilistic characteristics of PTTC using the established DDSPCE model. The comparison between DDSPCE and MCS is provided in Table \ref{tab:IEEE118_TTC_comp_dis}, in which $q = 0.0$ represents the continuous case (Scenario 1). The results show that a substantial reduction in the mean value of PTTC ($4.32\%$ when $q=0.1$, $9.14\%$ when $q=0.2$) and an increase in the variance ($12.84\%$ when $q = 0.1$, $25.65\%$ when $q = 0.2$) of PTTC after integrating the discrete random variables. These results evidently demonstrate the necessity and the pressing to incorporate discrete events in PTTC and ATC assessment to ensure the security of power grid, given the complex and aging transmission networks. 
\begin{table}[H]
\vspace{-0.3cm}
\setlength{\abovecaptionskip}{0.cm}
\caption{Comparison of the estimated statistics of the overall TTC by the MCS and DDSPCE methods for mixed case}
\renewcommand{\arraystretch}{1.2}
\label{tab:IEEE118_TTC_comp_dis}
\centering
\setlength{\tabcolsep}{1.6mm}{
\begin{tabular}{c|c|c|c|c|c|c}
\hline
\multirow{2}{*}{$q$} & \multicolumn{2}{c|}{MCS} & \multicolumn{4}{c}{DDSPCE}                             \\ \cline{2-7} 
                   & $\mu$          & $\sigma$      & $\mu$ &  $\frac{\Delta \mu}{\mu_{MCS}}\%$ & $\sigma$ & \multicolumn{1}{c}{$\frac{\Delta \sigma}{\sigma_{MCS}}\%$} \\ \hline
0.0                & 140.5468    & 29.6503    & 139.8183  & -0.5183 & 29.7303 &  0.2698\\ \hline                    
0.1                & 134.4748    & 33.4574    & 133.8207  & -0.4864 & 32.8810 & -1.7228                                \\ \hline
0.2                & 127.6986    & 37.2558   & 127.1441  & -0.4342 & 34.9234 & -6.2605                                 \\ \hline
\end{tabular}}
\vspace{-0.2cm}
\end{table}
To verify the performance of the proposed method, Fig. \ref{fig:TTC_comp_dis_2} presents the comparison between DDSPCE and MCS for $q = 0.1$. 
It is clear that the proposed method can achieve a good estimate for the mixed random inputs with much less computational time (about $\frac{1}{18}$ of the time required by MCS). Similar results can be achieved when $q=0.2$. 
\begin{figure}[H]
\vspace{-0.3cm}
\setlength{\abovecaptionskip}{-0.cm}
\setlength{\belowcaptionskip}{-0.9cm}
\centering
\includegraphics[width=0.36\textwidth]{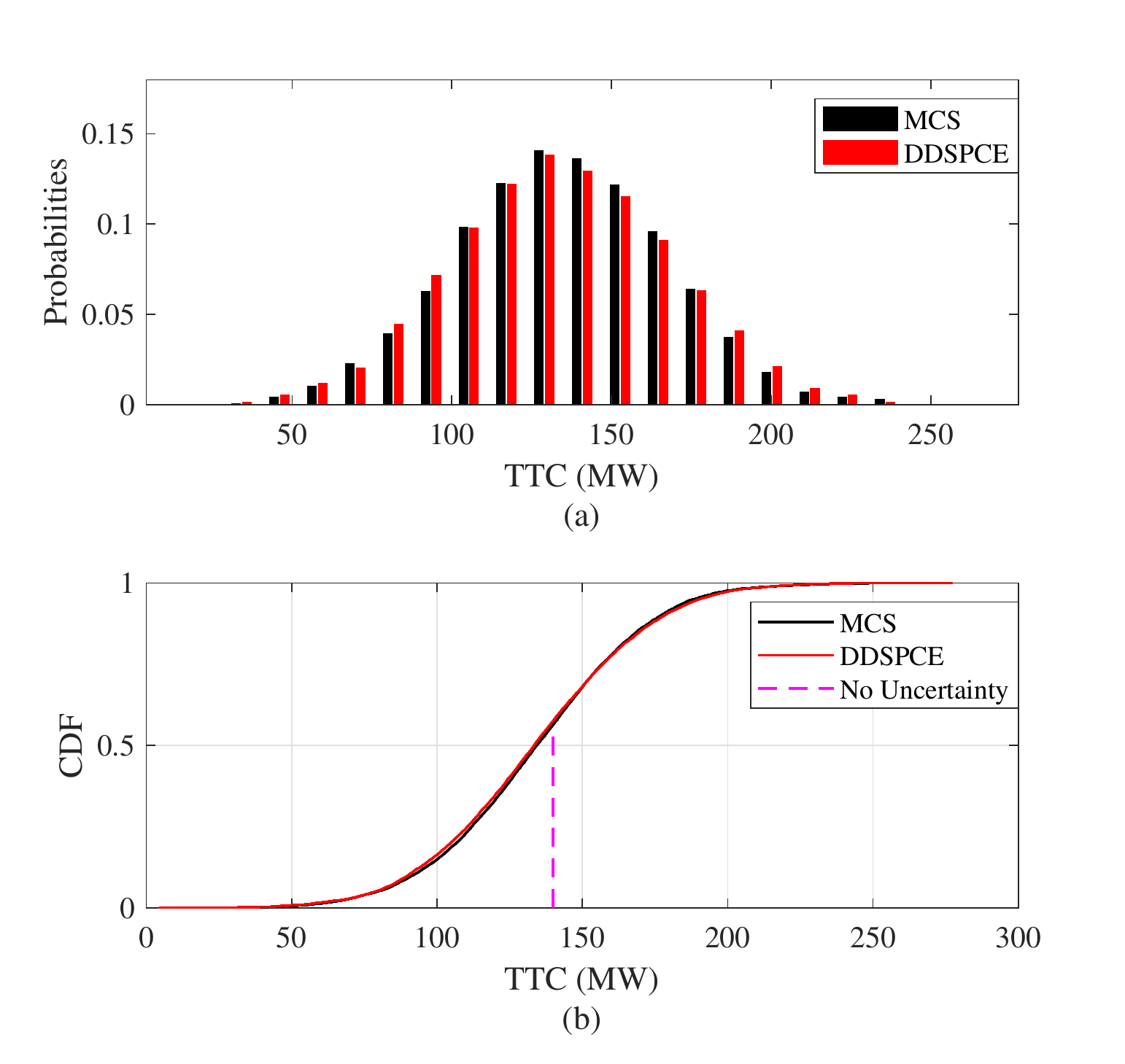}
\caption{The PDF and CDF of the PTTC calculated by the MCS and DDSPCE with probability $q = 0.1$. They are almost overlapped. The TRM for 95\% confidence level is 56.4333 MW and the corresponding ATC is 78.0415 MW.} 
\label{fig:TTC_comp_dis_2}
\end{figure}

Based on the probabilistic of PTTC, the TRM and the corresponding ATC can be calculated. 
Table \ref{tab:IEEE118_confid_levels_dis} shows the TRM and resulting ATC value at 95\%  confidence level. It can be clearly seen that the discrete random variables greatly reduce the ATC level ($14.09\%$ when $q=0.1$), indicating the great significance of accounting for discrete events in ATC assessment.
\begin{table}[H]
\vspace{-0.3cm}
\setlength{\abovecaptionskip}{0.cm}
\renewcommand{\arraystretch}{1.2}
\caption{The estimated TRM and resulting ATC (MW) for confidence level at 95\% based on the DDSPCE model}
\label{tab:IEEE118_confid_levels_dis}
\centering
\begin{tabular}{c|c|c|c}
\hline
$q$ & $\mathbb{E}($TTC$)$ (MW) & {TRM} (MW) & {ATC} (MW) \\
\hline
$0.0$ & 140.5468 & 49.7043 & 90.8425  \\
\hline
$0.1$ & 134.4748 & 56.4333 & 78.0415  \\
\hline
$0.2$ & 127.6986 & 61.3488 & 66.3498  \\
\hline
\end{tabular}
\end{table}

\subsection{Scenario 3: Considering $N-2$ Contingency and Adjustable Transformer} \label{section_scenario3} 
The third scenario aims to show that $N-2$ contingency as well as control devices such as adjustable transformers (e.g., Under-Load Tap Changer (ULTC) transformer) can be readily incorporated in the proposed PTTC formulation and ATC assessment. In the case studies thereafter, the configuration for continuous random variables and the transaction under study are the same as those in Scenario 1. Other parameters can be found in the aforementioned GitHub link.

\textit {1) With $N-2$ Contingency Considered} 

In this case, the contingency list contains five $N-2$ outages, which is presented in Table \ref{tab:IEEE118_TTC_N-2_list}. First, the deterministic TTC under the transaction considering $N-2$ contingency is determined as $88.8$ MW, much lower than 139.9  MW in Scenario 1, where only $N-1$ contingency is considered.  
\begin{table}[H]
\setlength{\abovecaptionskip}{0.cm}
\vspace{-0.4cm}
\renewcommand{\arraystretch}{1.3}
\caption{$N-2$ contingency list}
\label{tab:IEEE118_TTC_N-2_list}
\centering
\begin{tabular}{c|c|c|c|c|c}
\hline
\begin{tabular}[c]{@{}c@{}}Case No.\end{tabular} &  1                                                       & 2                                                        & 3                                                         &  4                                                       & 5                                                       \\ \hline
 Outages                                                 & \begin{tabular}[c]{@{}c@{}} L88-89\\  L89-92\end{tabular} & \begin{tabular}[c]{@{}c@{}}  L7-12\\  L12-16\end{tabular} & \begin{tabular}[c]{@{}c@{}} L13-15\\  L94-96\end{tabular} & \begin{tabular}[c]{@{}c@{}}  L49-54\\  L42-49\end{tabular} & \begin{tabular}[c]{@{}c@{}}  L91-92\\  L92-93\end{tabular} \\ \hline
\end{tabular}
\vspace{-0.2cm}
\end{table}

Next, the proposed DDSPCE model (\ref{eq:PCE}) is built to calculate the probabilistic characteristic of PTTC following the same procedures as the above scenarios (i.e., $H=2$, $M_p = 556$, $M_s =10000$). The estimated mean and standard deviation as well as their normalized errors in $\%$ computed by MCS, DDPCE and SPCE methods are provided in Table \ref{tab:IEEE118_TTC_N-2_results}. It can be seen that, when $N-2$ contingency is considered, the DDSPCE model can accurately estimate the statistics of the PTTC. Furthermore, the computational time of the DDSPCE model is only about $\frac{1}{18}$ of the time required by MCS. Similarly, it is believed that $N-K$ (${K}>2$) contingency can also be considered, if needed, by the proposed PTTC formulation, while the DDSPCE method can provide accurate estimation for the probabilistic characteristics of PTTC efficiently.  
\begin{table}[H]
\vspace{-0.3cm}
\setlength{\abovecaptionskip}{0.cm}
\renewcommand{\arraystretch}{1.3}
\caption{Comparison of the estimated statistics of the overall TTC with $N-2$ contingency considered by the  MCS, DDSPCE and SPCE methods}
\label{tab:IEEE118_TTC_N-2_results}
\centering
\begin{tabular}{c|c|c|c|c}
\hline
 Method &   $\mu$ &   $\sigma$ &   $\frac{\Delta\mu}{\mu_{MCS}}\%$ &   {$\frac{\Delta\sigma}{\sigma_{MCS}}\%$} \\
\hline
   MCS     &  90.2270  &   23.3703    &   -- &   --  \\
\hline
  DDSPCE  &   89.6840  &   23.4123    &   -0.6018   &   0.1797 \\
\hline
  SPCE   &  90.3824  &   22.8604    &    0.1722   &   -2.1818 \\
\hline 
\end{tabular}
\end{table}

\textit {2) With ULTC Transformer} 

In this case, the transformer from Bus 81 to Bus 80 (with fixed ratio 0.935 in pu) is replaced by a ULTC transformer to maintain the voltage magnitude within a range. The lower limit of the tap ratio is $0.775$ in pu and the upper limit of the tap ratio is set to be $1.185$ in pu. 

\begin{figure}[H]
\vspace{-0.4cm}
\setlength{\abovecaptionskip}{-0.2cm}
\setlength{\belowcaptionskip}{-0.9cm}
\centering
\includegraphics[width=0.36\textwidth]{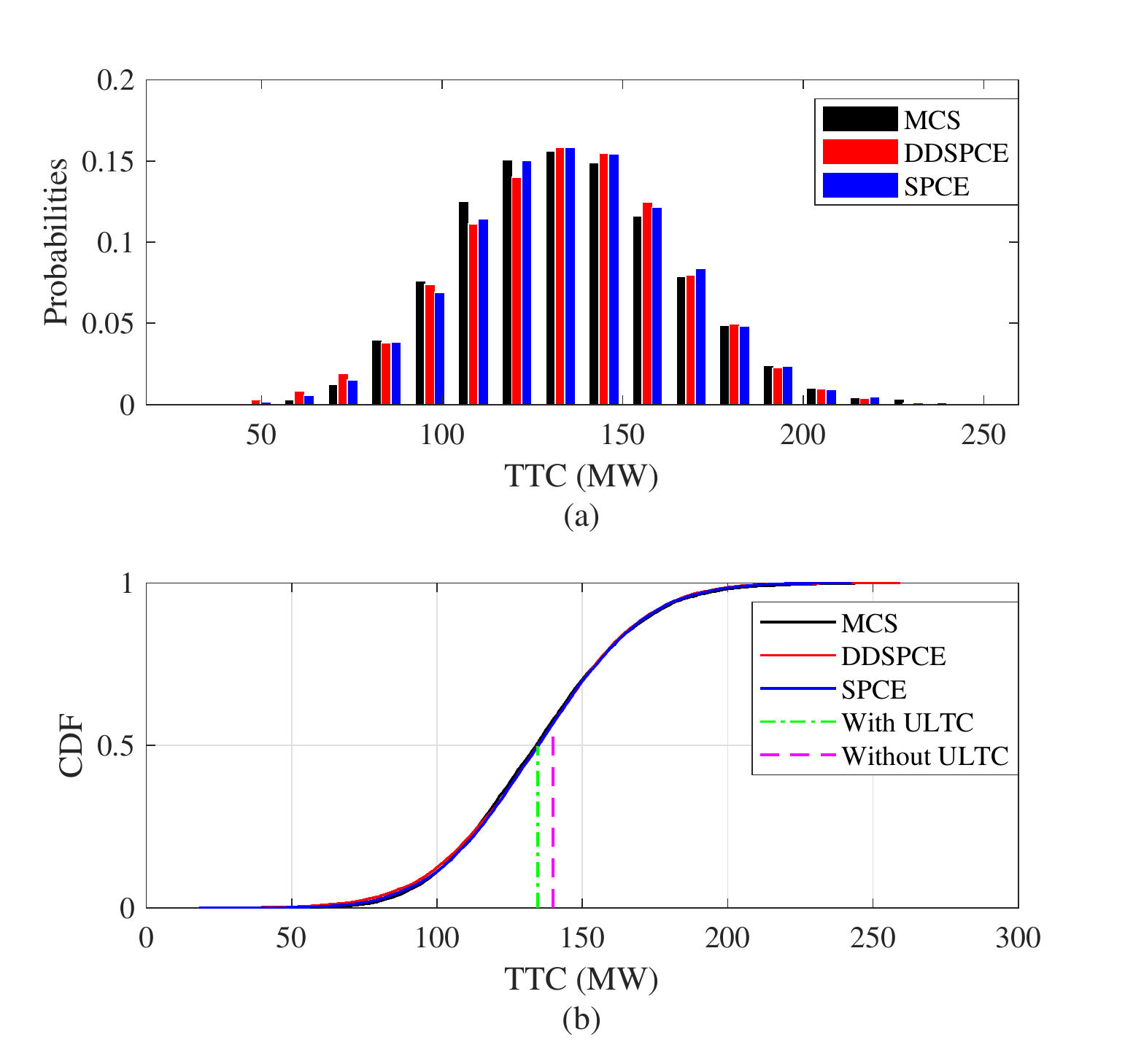}
\caption{The PDF and CDF of the PTTC calculated by the MCS, DDSPCE and SPCE  with ULTC transformer at Bus 81-Bus 80. 
They are almost overlapped. The TRM for 95\% confidence level is 49.4894 MW and the corresponding ATC is 85.6214 MW. The deterministic TTC considering ULTC (the green dash-dot line) is slightly less than the deterministic TTC without ULTC (the magenta dashed line) obtained in Scenario 1.} 
\label{fig:TTC_comp_ultc}
\vspace{-0.2cm}
\end{figure}

The deterministic TTC value is 134.6 MW  calculated by the DSATools/VSAT solver. A $5.3$ MW decrease is observed after integrating the ULTC compared to the deterministic TTC value 139.9 MW in Scenario 1 (see Fig.\ref{fig:TTC_comp_ultc} (b)). Such decrease in the TTC is expected as a compromise of maintaining voltage level. Next, the proposed DDSPCE model is exploited to calculate the statistics of PTTC similar to the previous scenarios (i.e., $H=2$, $M_p = 556$, $M_s =10000$). 
The estimation results and the comparisons between DDSPCE, SPCE and MCS are summarized in Table \ref{tab:IEEE118_TTC_ultc}, 
showing that, when ULTC is incorporated, the DDSPCE method can still provide accurate estimation for the probabilistic characteristics of PTTC. Similar efficiency  to the previous scenarios ($\approx\frac{1}{18}$ of time consumed by MCS) is also observed in this case. 
\begin{table}[H]
\vspace{-0.3cm}
\setlength{\abovecaptionskip}{0.cm}
\renewcommand{\arraystretch}{1.3}
\caption{Comparison of the estimated statistics of the overall TTC with ULTC considered by the MCS, DDSPCE and SPCE methods}
\label{tab:IEEE118_TTC_ultc}
\centering
\begin{tabular}{c|c|c|c|c}
\hline
 Method &   $\mu$ &   $\sigma$ &   $\frac{\Delta\mu}{\mu_{MCS}}\%$ &   {$\frac{\Delta\sigma}{\sigma_{MCS}}\%$} \\
\hline
   MCS     &  135.1108  &   29.4460    &   -- &   --  \\
\hline
  DDSPCE  &   134.4281  &   29.5267    &   -0.5053   &   0.2741 \\
\hline
  SPCE   &  135.2722  &  29.2345    &    0.1195   &   -0.7183 \\
\hline 
\end{tabular}
\vspace{-0.2cm}
\end{table}

Likewise, the TRM and resulting ATC can be determined from the calculated probabilistic characteristics of PTTC. Table \ref{tab:IEEE118_confid_levels_ultc} shows the comparison of the estimated TRM and the resulting ATC at 95\% confidence level with and without ULTC. It can be seen that these two cases share a similar TRM value when the same random inputs are considered. However, a lower mean TTC obtained after integrating the ULTC results in a slightly lower ATC in this case. 
\begin{table}[]
\setlength{\abovecaptionskip}{0.cm}
\setlength{\belowcaptionskip}{-0.8cm}
\renewcommand{\arraystretch}{1.2}
\caption{Comparison of the estimated TRM and resulting ATC at 95\%  confidence level with and without ULTC by the DDSPCE model}
\label{tab:IEEE118_confid_levels_ultc}
\centering
\begin{tabular}{c|c|c|c}
\hline
 Control device & $  \mathbb{E}( $TTC$)$ (MW) & {TRM} (MW) &  {ATC} (MW) \\
\hline
 Without ULTC &  140.5468 &  49.7043 &  90.8425  \\
\hline
 With ULTC &  135.1108 &  49.4894 &  85.6214  \\
\hline
\end{tabular}
\begin{tablenotes}
\item * The results of without ULTC is obtained directly from Scenario 1.
\end{tablenotes}
\vspace{-0.3cm}
\end{table}

\section{Conclusion and Perspectives\color{black}}
In this paper, we have proposed a novel data-driven sparse PCE  method to assess the probabilistic characteristics of PTTC without requiring pre-assumed probability distributions of random inputs (e.g., RES, load variation and line outages). The proposed DDSPCE method,  exploiting data sets directly, can handle a large number of mixed (both continuous and discrete) random inputs. 
In addition, the sparse PCE scheme has been integrated to reduce computational effort.  
Simulation results reveal that the proposed DDSPCE method can obtain accurate estimation for the probabilistic characteristics of PTTC with high efficiency. Moreover, the necessity of incorporating discrete random variables in PTTC and ATC assessment has been demonstrated by seeing a large impact on the statistics of PTTC and a significant decrease in the ATC level after integrating the discrete random inputs.  In the near future, we will exploit advanced dimension reduction techniques to further reduce the growing computational efforts as the number of random inputs increases. In addition, global sensitivity analysis will be leveraged to  reduce the dimension of random input variables and thus to improve the computational efficiency for PTTC and ATC assessment. 



%




%

\ifCLASSOPTIONcaptionsoff
  \newpage
\fi

\begin{IEEEbiography}[{\includegraphics[width=1in,height=1.25in,clip,keepaspectratio]{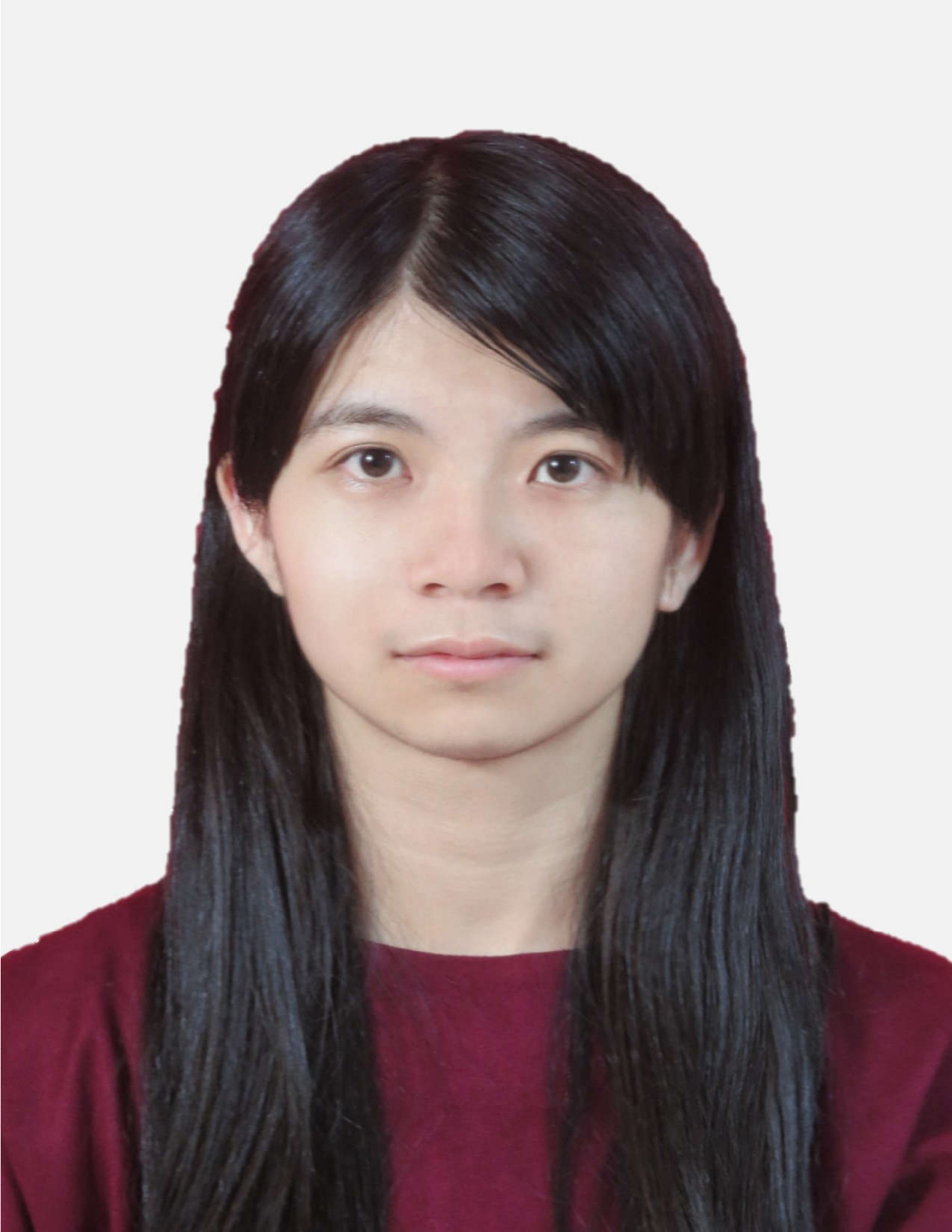}}]{Xiaoting Wang}
is currently a Ph.D. student in the Department of Electrical and Computer Engineering at McGill University, Montreal, QC, Canada. She received the M.S. degree in Control Science and Engineering from Harbin Institute of Technology, Shenzhen, China, in 2019 and the B.S. degree in Electrical Engineering and Automation from Fuzhou University, Fuzhou, China, in 2016. Her research interests include the uncertainty quantification of power system stability and security, fast power system stability enhancement methods.
\end{IEEEbiography} 

\begin{IEEEbiography}[{\includegraphics[width=1in,height=1.25in,clip,keepaspectratio]{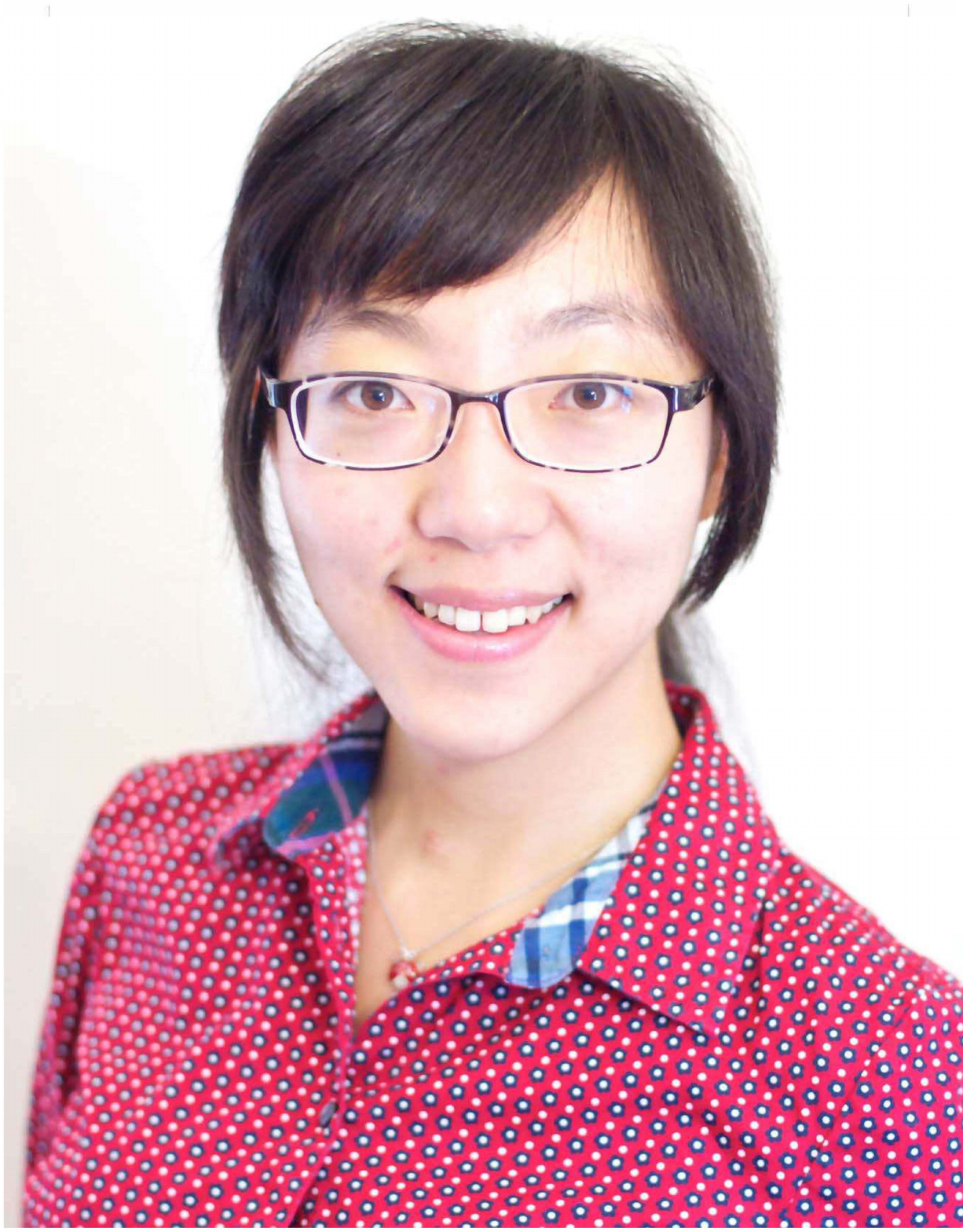}}]{Xiaozhe Wang}
is currently an Assistant Professor in the Department of Electrical and Computer Engineering at McGill University, Montreal, QC, Canada. She received the Ph.D. degree in the School of Electrical and Computer Engineering from Cornell University, Ithaca, NY, USA, in 2015. Her research interests are in the general areas of power system stability and control, uncertainty quantification in power system security and stability, and wide-area measurement system (WAMS)-based detection, estimation, and control. She is an IEEE senior member, serving on the editorial boards of IEEE Transactions on Power Systems, Power Engineering Letters, and IET Generation, Transmission and Distribution.
\end{IEEEbiography}

\begin{IEEEbiography}[{\includegraphics[width=1in,height=1.25in,clip,keepaspectratio]{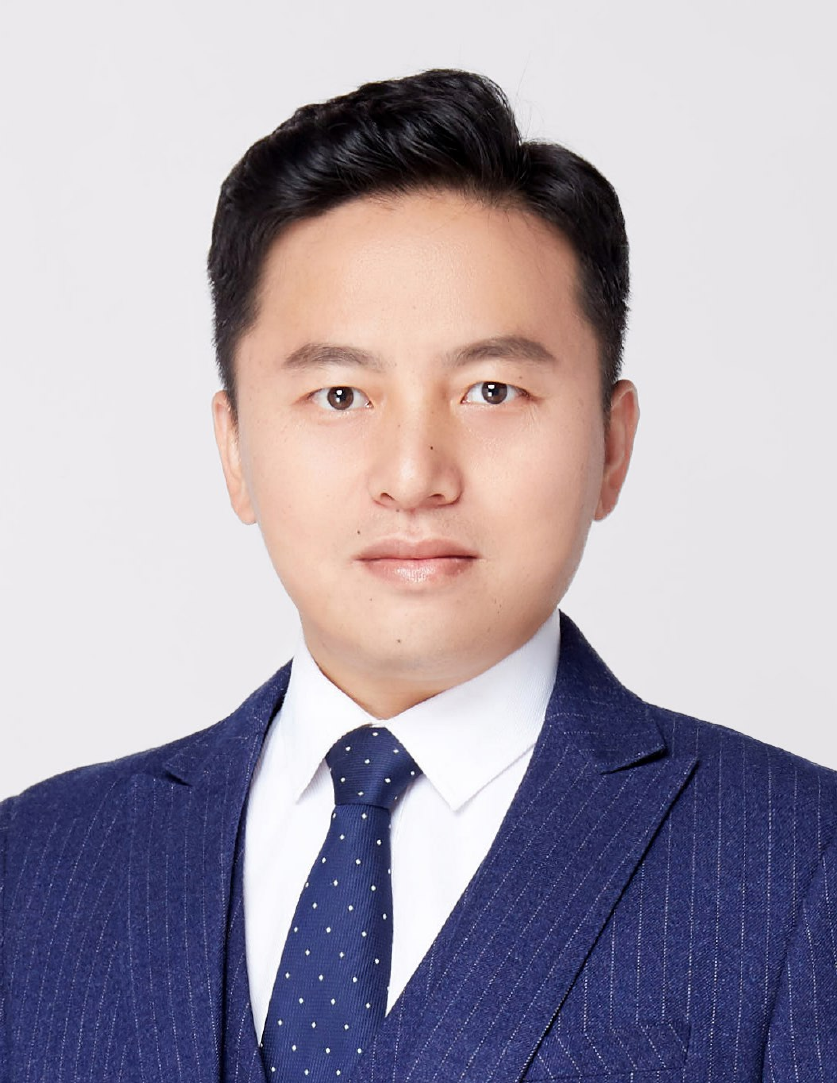}}]{Hao Sheng} 
received the B.S. degree from North China Electric Power University, Baoding, China, in 2003, the M.S. degree from Northeast Electric Power University, Jilin, China, in 2007, and the Ph.D. degree from the School of Electrical and Automation Engineering, Tianjin University, Tianjin, China, in 2014, all in electrical engineering.
From 2007 to 2012, he was an software engineer with the R\&D Centre, Beijing Sifang Automation Co., Ltd., Beijing, China. From 2014 to 2017, he was a Postdoctoral Fellow with the School of Electrical and Computer Engineering, Cornell University, Ithaca, NY, USA. From 2017 to 2019, he was a Postdoctoral Fellow with the Department of Electrical and Computer Engineering, McGill University, Montreal, QC, Canada.
He is currently an Associate Professor with the School of Electrical and Information Engineering, Tianjin University, China. His research interests include power system stability and control, power system planning and reliability, and probabilistic security assessment and optimization.
\end{IEEEbiography}

\begin{IEEEbiography}[{\includegraphics[width=1in,height=1.25in,clip,keepaspectratio]{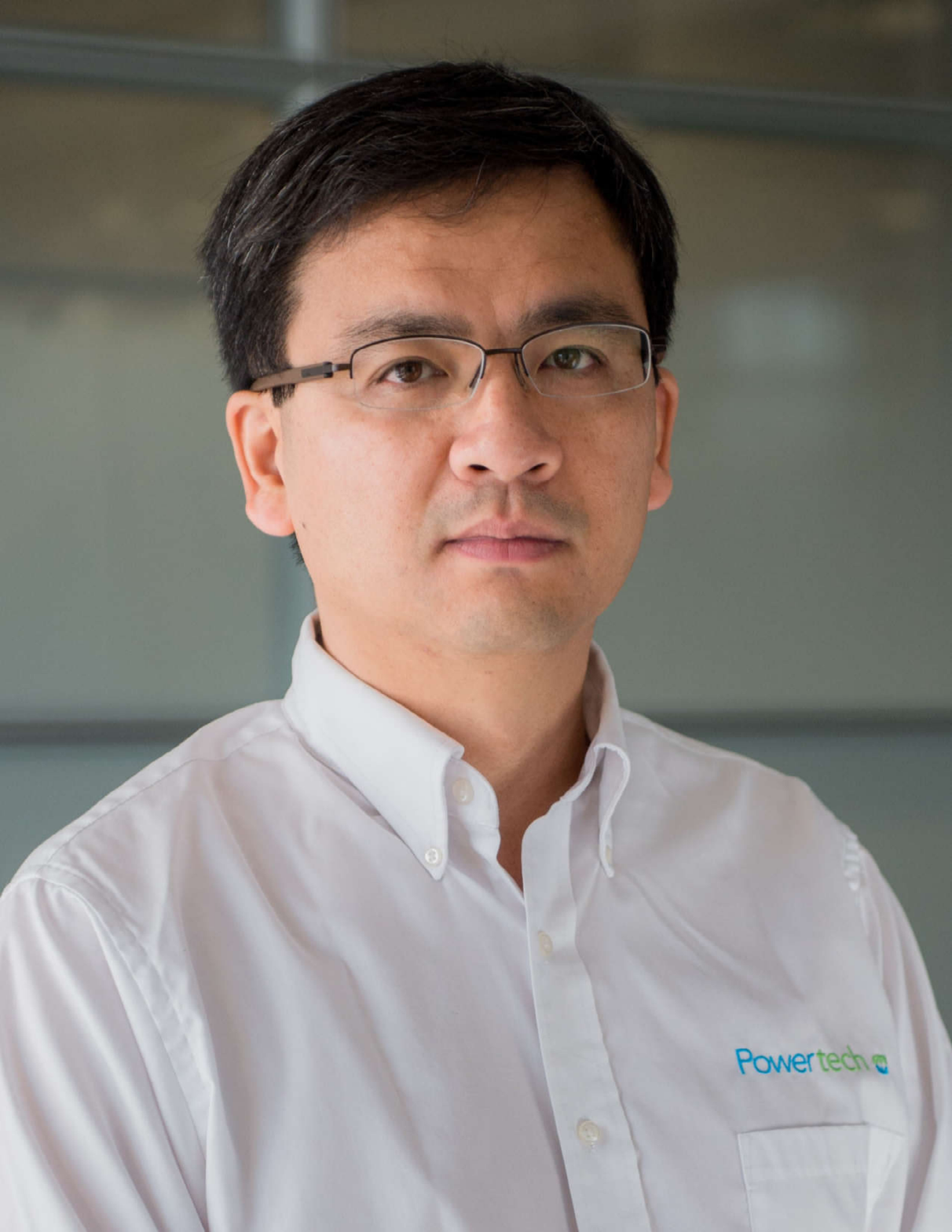}}]{Xi Lin}
received the Bachelor degree in mechanical engineering from the Tsinghua University, China in 1997 and the Master degree in electrical engineering from the Nanjing Automation Research Institute (NARI), China, in 2000. He received the Ph.D. degree in electrical engineering from the University of Manitoba, Canada in 2011. He has been with Powertech Labs Inc. in Surrey, Canada since 2007. He is a member of the development team of the DSATools software. His research interests include power system stability and control, power system simulation. Xi is an IEEE Senior Member.
\end{IEEEbiography} 







\end{document}